\documentclass[lettersize,journa]{IEEEtran}
\IEEEoverridecommandlockouts
\usepackage{cite}
\usepackage{float}
\usepackage{wrapfig}
\usepackage{amsmath,amssymb,amsfonts}
\usepackage{algorithmic}
\usepackage{graphicx}
\usepackage{subfig}
\usepackage{textcomp}
\usepackage{wrapfig}
\usepackage{xcolor}
\def\BibTeX{{\rm B\kern-.05em{\sc i\kern-.025em b}\kern-.08em
    T\kern-.1667em\lower.7ex\hbox{E}\kern-.125emX}}
\begin{document}
\title{Wireless Connectivity and Localization for Advanced Air Mobility Services}


\author{
\IEEEauthorblockN{Priyanka Sinha, Md Moin Uddin Chowdhury, Ismail Guvenc, David W. Matolak, and Kamesh Namuduri}
\IEEEauthorblockN{
\thanks{This manuscript was submitted for review on Feb. 15, 2022. This work has been supported in part by NASA and NSF through the awards NX17AJ94A, CNS-1814727, CNS-1453678, and CNS-1910153.
\\
P. Sinha is with North Carolina State University, Raleigh, NC 27606 USA (e-mail: psinha2@ncsu.edu).
\\
Md M. U. Chowdhury is with North Carolina State University, Raleigh, NC 27606 USA (e-mail: mchowdh@ncsu.edu).
\\
I. Guvenc is with North Carolina State University, Raleigh, NC 27606 USA (e-mail: iguvenc@ncsu.edu).
\\
D. W. Matolak is with University of South Carolina, Columbia, SC 29208 USA (e-mail: matolak@cec.sc.edu).
\\
K. Namuduri is with University of North Texas, Denton, TX 76203 USA (e-mail: kamesh.namuduri@unt.edu).} 
}}

\maketitle

\begin{abstract}
By serving as an analog to traffic signal lights, communication signaling for drone to drone communications holds the key to the success of advanced air mobility (AAM) in both urban and rural settings. Deployment of AAM applications such as air taxis and air ambulances, especially at large-scale, requires a reliable channel for a point-to-point and broadcast communication between two or more aircraft.  Achieving such high reliability, in a highly mobile environment, requires communication systems designed for agility and efficiency. This paper presents the foundations for establishing and maintaining a reliable communication channel among multiple aircraft in unique AAM settings. Subsequently, it presents concepts and results on wireless coverage and mobility for AAM services using cellular networks as a ground network infrastructure. Finally, we analyze the wireless localization performance at 3D AAM corridors when cellular networks are utilized, considering different corridor heights and base station densities. We highlight future research directions and open problems to improve wireless coverage and localization throughout the manuscript.      
\end{abstract}
\begin{IEEEkeywords}
5G, advanced aerial mobility (AAM), air corridors, drones, handover, localization, unmanned aerial systems (UAS), urban air mobility (UAM).
\end{IEEEkeywords}

\section{Introduction}

Unmanned aircraft systems (UAS) deployed for commercial applications such as air taxis and air ambulances are expected fly between 1000 feet to 2000 feet above  ground level. Supporting such UAS applications requires dedicated air corridors for safe and efficient operations of aerial vehicles. To this end, the National Aeronautics and Space Administration (NASA) has been recently working on the advanced air mobility (AAM) mission, which will \emph{``help emerging aviation markets to safely develop an air transportation system that moves people and cargo between places previously not served or underserved by aviation''}~\cite{NASA_AAM_Vision}. 

Successful design and operation of AAM services require reliable air-to-air (AA) and air-to-ground (AG) wireless connectivity in air corridors. Cellular networks, due to their widely deployed base station (BS) infrastructure, are the prime candidates for providing wireless connectivity services to aerial vehicles (AVs)~\cite{bauranov2021designing}. However, cellular networks are not originally designed for serving aerial mobile equipment. For example, their antennas are down-tilted to provide optimum coverage to mobile equipment on the ground, and aerial coverage, especially at high altitudes, can be unreliable and patchy~\cite{lin2018sky,chowdhury2021ensuring}. Moreover, there can be significant interference from the aerial mobile equipment to the uplink transmissions of cellular users on the ground, as the AVs tend to have line-of-sight (LOS) with a large number of cellular base stations operating in the same spectrum~\cite{amorim2018measured,al2017modeling}. Therefore, use of cellular networks for AAM services require careful design and optimization for maintaining reliable air corridor coverage while not also disrupting existing services. 

Although UAS will communicate over AG links for multiple purposes, AA links will also be required for several functions~\cite{muna2021air}. For example, during close encounters near air corridor intersections, AVs may need to negotiate their maneuvers through sharing their intent with one another.  Such information exchange in real-time requires a reliable communication channel between the two aircraft involved in the negotiation. The AVs will also need reliable AA links for surveillance purpose, e.g. for detect-and-avoid (DAA) of nearby AVs, to ensure safety of flight. Other applications for AA links include information relay (e.g., one drone informing others of airspace status), sharing common mission data, and emergency communications in case one or more AG links incurs an outage.

Finally, for ensuring safe and secure AAM operations, knowing the locations of AVs carries critical importance. Use of the global positioning services (GPS) technology is commonly used by the AV itself and the corresponding AV location is reported over the AG links. However, there may be cases where GPS location estimate is not reliable, or not even available, e.g. due to propagation effects in urban canyons, radio frequency interference (RFI), jamming, or failure of GPS equipment itself~\cite{ellis2020time}, which can cause serious threats for safe AAM operations. Cellular networks, while providing wireless coverage at air corridors, can also be simultaneously utilized for localizing and tracking the AVs, in cases where GPS information is not reliable or available.  

Considering all these challenges, this paper outlines the current status and future directions in providing reliable wireless connectivity and localization for AVs to operate safely in air corridors. First, in Section~II, we  present the concept of air corridors  and the rules of engagement for AVs flying in air corridors. 
In Section~III, we analyze wireless coverage for AAM services when a cellular ground infrastructure is used. Both down-tilted (legacy) and up-tilted antennas are considered at the cellular BSs, while considering different corridor heights and BS densities. Section~IV studies the mobility and handover characteristics of AVs when they travel across a terrestrial network. In Section~V, localization accuracy of a cellular network is evaluated when the BSs and the AV use multiple antennas to improve accuracy. Finally, Section~VI concludes the paper.

\begin{figure}
\begin{center}
    \includegraphics[width=\linewidth]{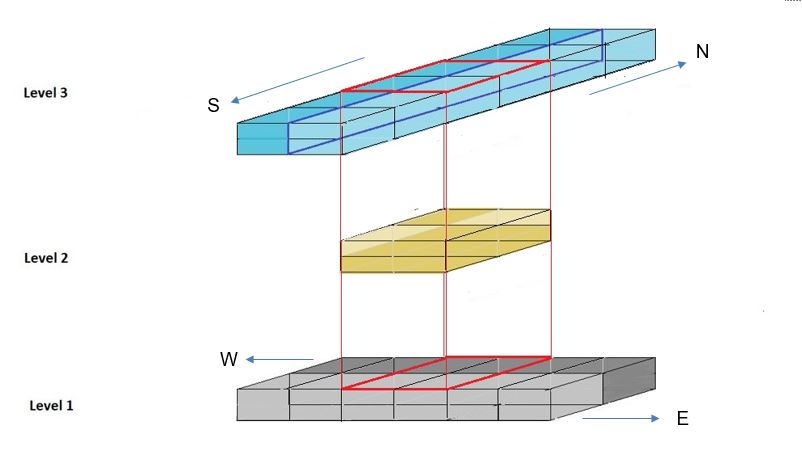}
\end{center}
\caption{Multi-level air corridor concept for advanced air mobility \cite{muna2021air}.}\label{Fig:CorridorLevels}
\end{figure}

\section{Overview of Air Corridor Concept}
Air corridors will be an integral part of AAM infrastructures. This paper assumes the use of a three dimensional air corridor model as in Fig.~1, in which the airspace is divided into two layers throughout the airspace, except at intersections. The top layer of the air corridor accommodates southbound and northbound traffic whereas the bottom layer accommodates eastbound and westbound traffic. At intersections, there is also a middle layer which is used by the vehicles for hovering when changing directions. Each layer may accommodate more than one lane (also called skylanes). We also point out that -- just as in current civil aviation with on-board pilots -- the actual position of these lanes may change if needs arise, due to traffic density, natural events, among other possible reasons.

\subsection{Rules of Engagement}

We consider the following rules of engagement between AVs that use the air corridors. 
\begin{enumerate}
    \item 3D space is organized into uniformly spaced 3D rectangular prisms. At any given time, a specified minimal volume within each prism can be occupied by only one vehicle. The dimensions of each prism also depend on safety guidelines and vehicle class.
    \item A vehicle needs to make sure that the prism it is entering at time ($t+1$) is going to be empty at time ($t+1$). Emptiness must be quantitatively defined for given airspace locations/uses.
    \item Overtaking (passing) is not expected to occur in air corridors. If one vehicle slows down, the vehicles behind the slow vehicle also need to slow down. 
    \item Vehicles must keep a minimum safety distance (SD) among them.
\end{enumerate}
Enforcing these rules of engagement requires reliable vehicle-to-vehicle links,as will be discussed next.

\subsection{The Need for Vehicle to Vehicle Communications}
Assume that normal flight operations are taking place in air corridors. Three vehicles A, B, and C are flying in the same direction within a track (or skylane),  maintaining the minimum SD among them. Suddenly, vehicle A detects an obstacle (a dense cloud, for example, as shown in Fig.~2) and slows down or stops and hovers. As it may not be able to move further, vehicle A needs to share the information that it gathered about the airspace hazard with the vehicles (B and C) behind it, as well as with the ground control station (GCS); it may next need to seek a possible opening of an emergency lane. This use case scenario includes a combination of DAA movements and Vehicle to Vehicle Communications (V2V), here, AA communications. 

\begin{figure}[t!]
    \begin{center}
        \includegraphics[width=\linewidth]{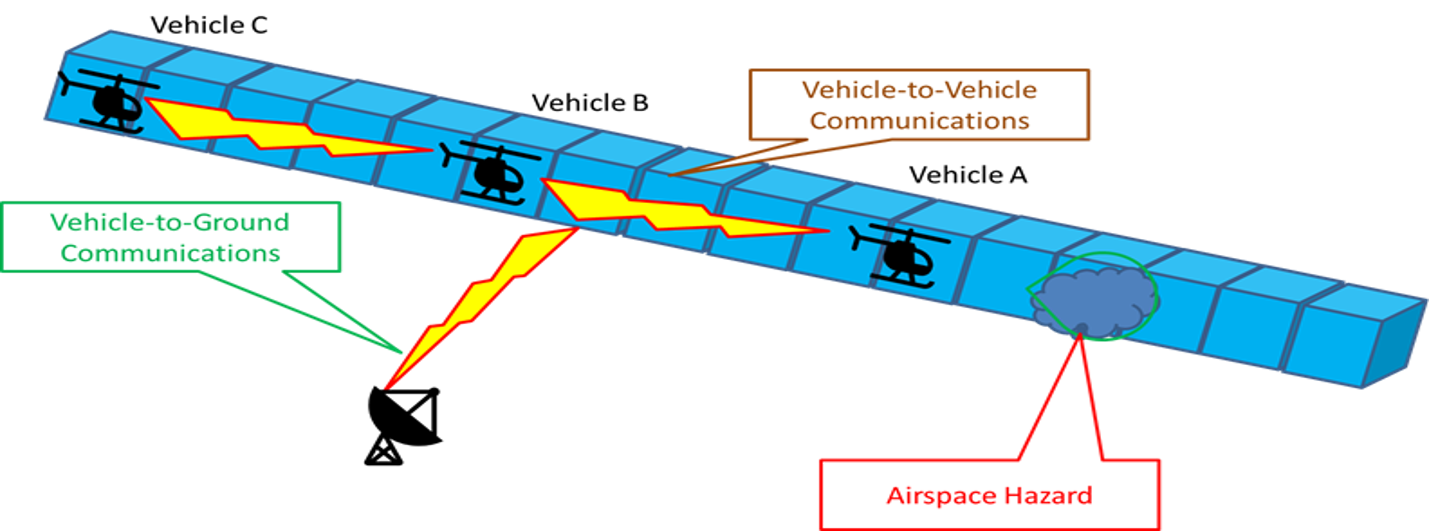}
    \end{center}
        \caption{Vehicle to vehicle communications for safe navigation in AAM corridors [Courtesy, Keven Gambold].}\label{Fig:CorridorHazard}
\end{figure}

Another situation that requires V2V communications is at intersections. An intersection can be three dimensional as shown in Fig.~1, or 2-dimensional as shown in Fig.~3, which depicts a circular intersection or a roundabout. Vehicles making turns at such roundabouts need to plan their 2D or 3D trajectories and maintain minimum SD among them. Tactical deconfliction in such close encounters requires direct V2V communications, in addition to potentially other means of communications, including satellite or ground-based communications; these latter approaches may not meet the latency requirements.

\begin{figure}[t!]
    \begin{center}
        \includegraphics[width=\linewidth]{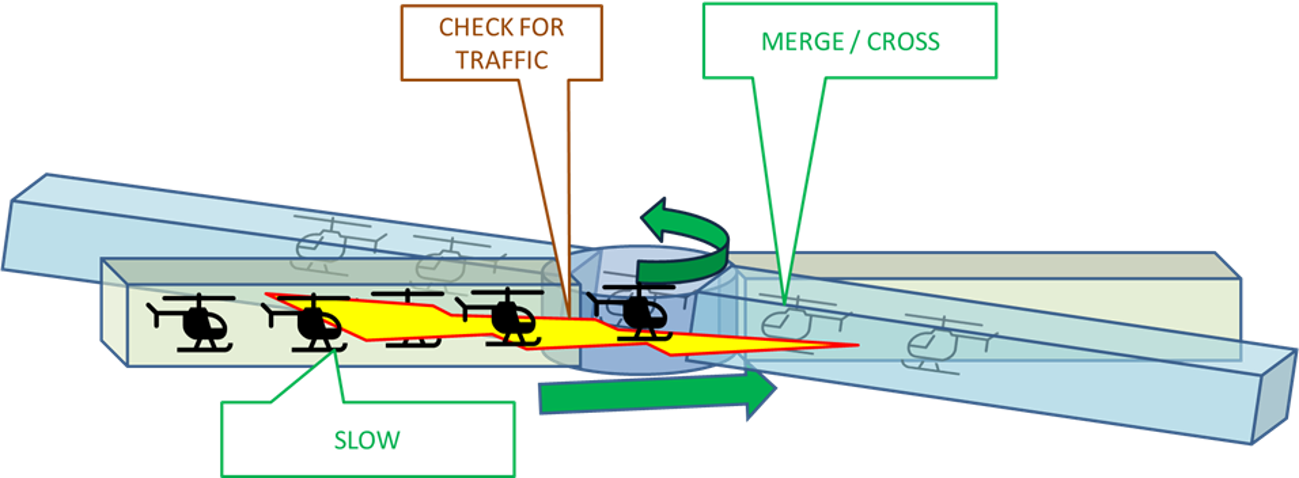}
    \end{center}
    \caption{Roundabouts in AAM corridors and V2V connectivity for vehicle coordination [Courtesy, Keven Gambold].}
    \label{Fig:CorridorIntersection}
\end{figure}

\subsection{Future Research Directions}
The concept of AAM corridors is still evolving. AAM corridors are expected to support vehicles of many classes. Reliable wireless connectivity in air corridors is a requirement for sharing situational awareness among the vehicles in real-time. In some regions, manned vehicles may need to cross AAM corridors. Hence, there is a need to design common communication strategies for both manned and unmanned AVs. Effective application of the V2V technology developed for 5G wireless services~\cite{8741719} needs to be further studied and evaluated in AAM air corridor scenarios.

\section{Wireless Coverage for Aerial Vehicles}

The AAM services will require diverse requirements such as seamless connectivity, high reliability, low latency, and continuous (high-availability) coverage across the drone corridors. In this section, after reviewing candidate wireless technologies for maintaining wireless connectivity for AAM, we study the use of cellular technology for providing wireless coverage at air corridors.  

\subsection{Communication Technologies for Safe and Reliable UAS}

The UAS flying autonomously through the drone corridors are required to maintain reliable and seamless command and control (C\&C) communication in the downlink with the deployment authority. Apart from this, the data rate and latency requirements for the AVs could vary depending on the UAS application and associated telemetry traffic~\cite{facom}. For instance, the data rate required will be relatively high for a live video stream via a UAV flying in the beyond visual line of sight (BVLOS) scenario. According to the Third Generation Partnership Project (3GPP), the reliability requirement of C\&C links is greater than $99.99$\%~and the latency requirement can vary between $50-100$~ms~\cite{3gpp}. Such stringent requirements could be met by several candidate technologies such as cellular networks or other dedicated terrestrial-based networks, satellite communication (satcom), quasi-stationary high altitude platforms (HAPs), or by combining two or more of these technologies~\cite{facom}. For satcom, several issues arise, including costly infrastructure and hardware, latency, and for low- or medium earth orbit (LEO, MEO) systems, high Doppler shifts and rates. For LEO systems, fast beam tracking for connecting between multiple low earth orbit satellites is needed~\cite{geraci2021}. Tracking with HAPS should be easier than with LEO/MEO satcom systems, but imperfect HAPs station-keeping can still shift antenna main beams from the intended direction, and hence, robust HAPS beam tracking will also be required to serve dynamic UAS.

\subsection{Air-to-Air Communication}

As noted earlier, for the safe and efficient implementation of UAS in the drone corridors, AA communication will play a critical role. The AVs in the UAS will need to interconnect with each other, and this will yield a special kind of mobile ad-hoc network where the movements of the AVs are largely predictable~\cite{facom}. This kind of direct communication between AVs can overcome the costly resource management, high Doppler, and longer round-trip delay challenges associated with satcom and HAPs, and could be a key enabler of large-scale city-wide UAS deployment. For DAA purposes, an AV can broadcast its presence to other nearby AVs to avoid unsafe situations in the drone corridors, essentially functioning like existing automatic dependent surveillance-broadcast (ADS-B) systems on civil passenger aircraft today. The current ADS-B system is unlikely to have sufficient capacity for many envisioned AAM settings and applications, hence augmentations to this system, or new approaches, must be employed.

In addition to AA communication for AV coordination, AVs can also relay data outside their coverage area to deliver time-sensitive data to other AVs or BSs; this can reduce backhaul traffic. As a result, the overall spectral and energy efficiency of the network can be improved. The exact number of AVs flying at any given time also affects AA communication requirements and performance. Moreover, the trajectories of the AVs (``intent") will also need to be shared for efficient routing of the information through AA communication.

In current air traffic management (ATM) systems, there is minimal support for V2V communications. For example, the traffic collision avoidance system (TCAS), which is designed to reduce the incidence of mid-air collision (MAC) between aircraft, alerts pilots of the presence of other transponder-equipped aircraft which may present a threat of a MAC. TCAS will issue traffic advisories (TA) and resolution advisories (RA), when appropriate. When an RA is issued to conflicting aircraft, the vehicles (pilots) are expected to perform the required actions (such as ascend and descend) and this process is manual. Such concepts can be extended to UAVs using V2V communications. Existing air traffic controller (ATC) to pilot communications in the aviation very high frequency (VHF) band are another example where AA communications takes place. Here, both aircraft and ATC essentially broadcast their transmissions, allowing nearby aircraft to receive and monitor.

\begin{figure*}[t]
\centering
	\subfloat[ISD = $1000$~m, $h_{\textrm{UAV}}$ = $100$~m]{
			\includegraphics[width=.25\linewidth]{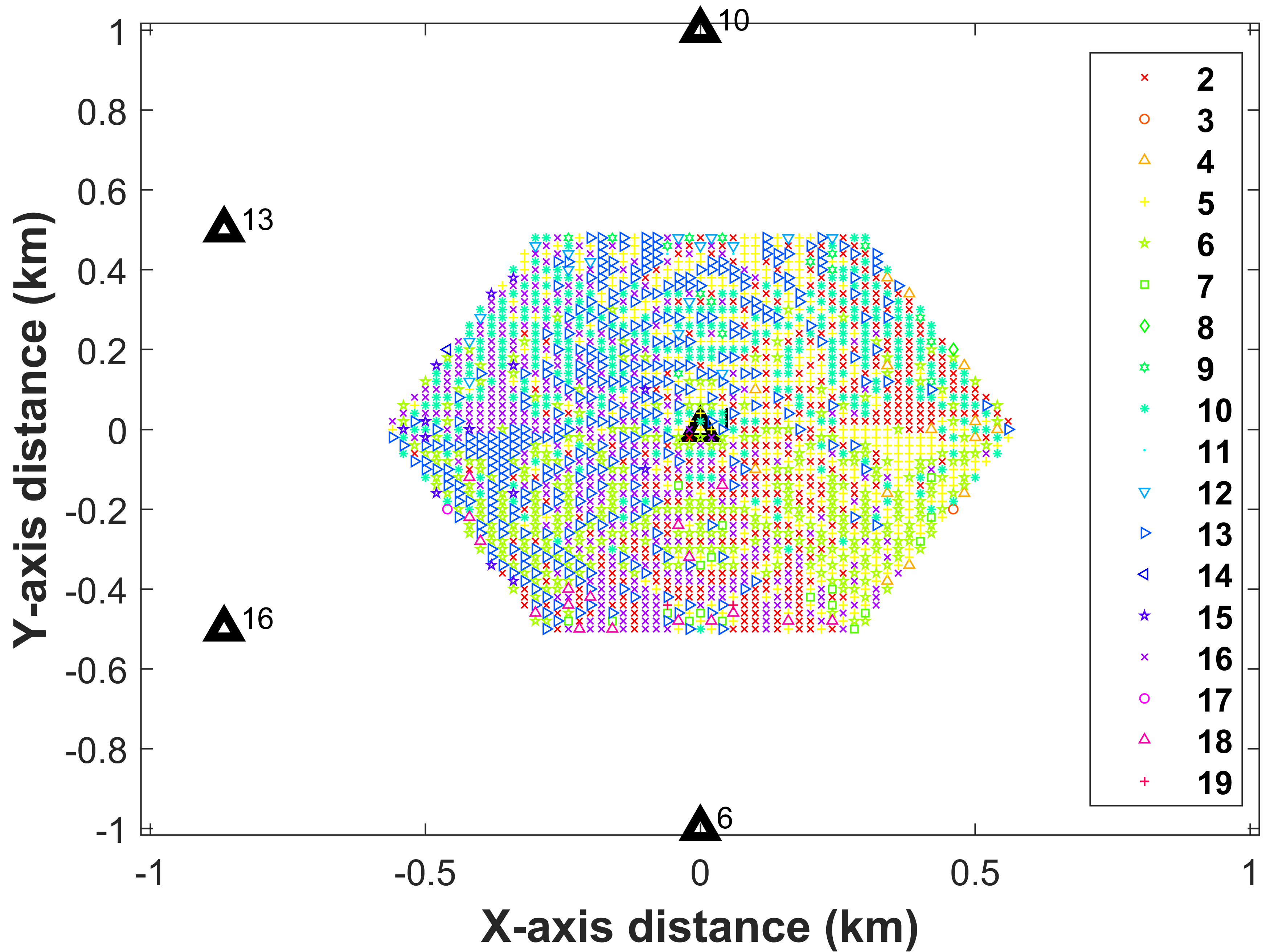}}
		\subfloat[ISD = $1000$~m, $h_{\textrm{UAV}}$ = $500$~m]{
			\includegraphics[width=.25\linewidth]{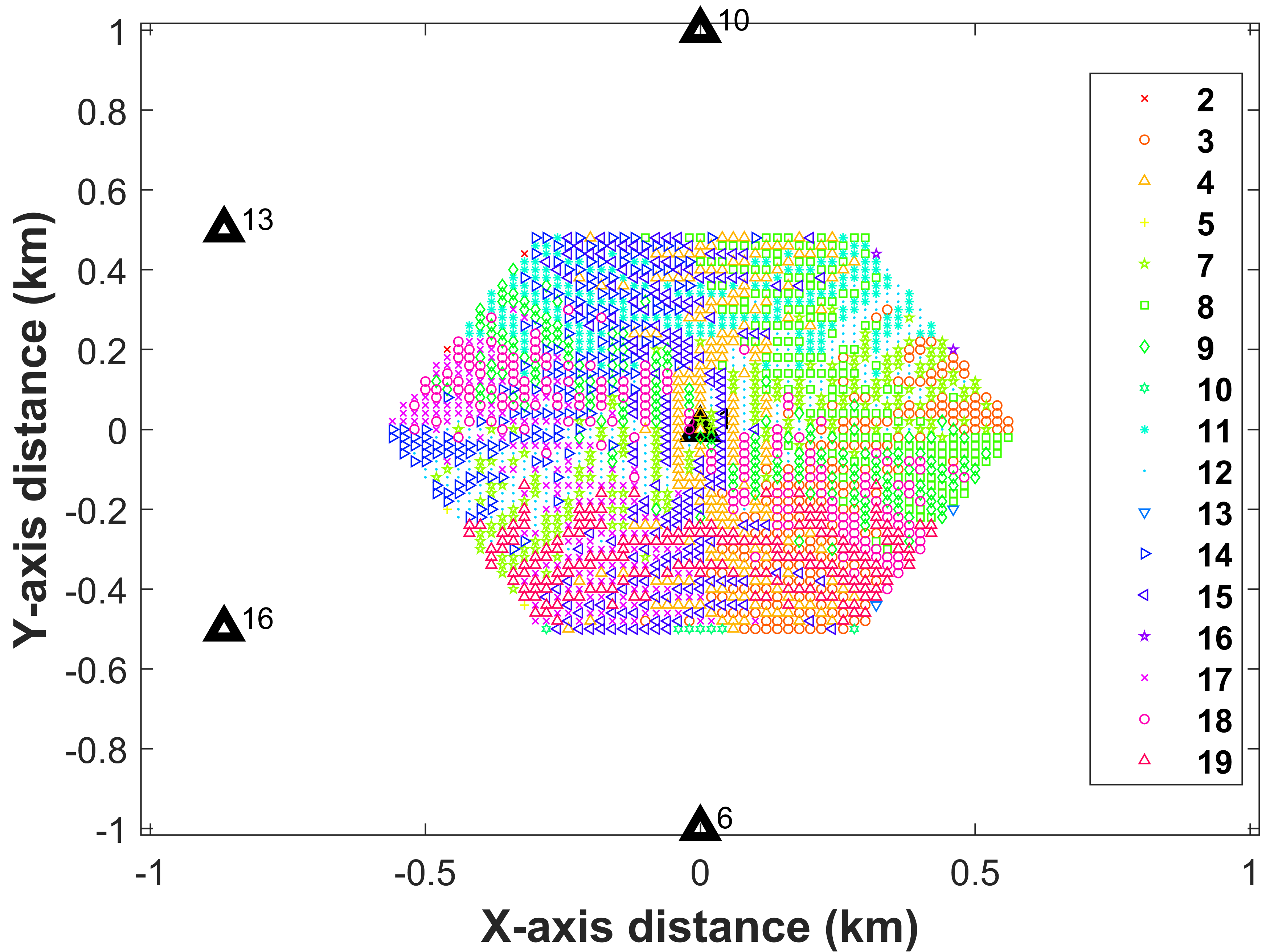}} 
			\subfloat[ISD = $2000$~m, $h_{\textrm{UAV}}$ = $100$~m]{
			\includegraphics[width=.25\linewidth]{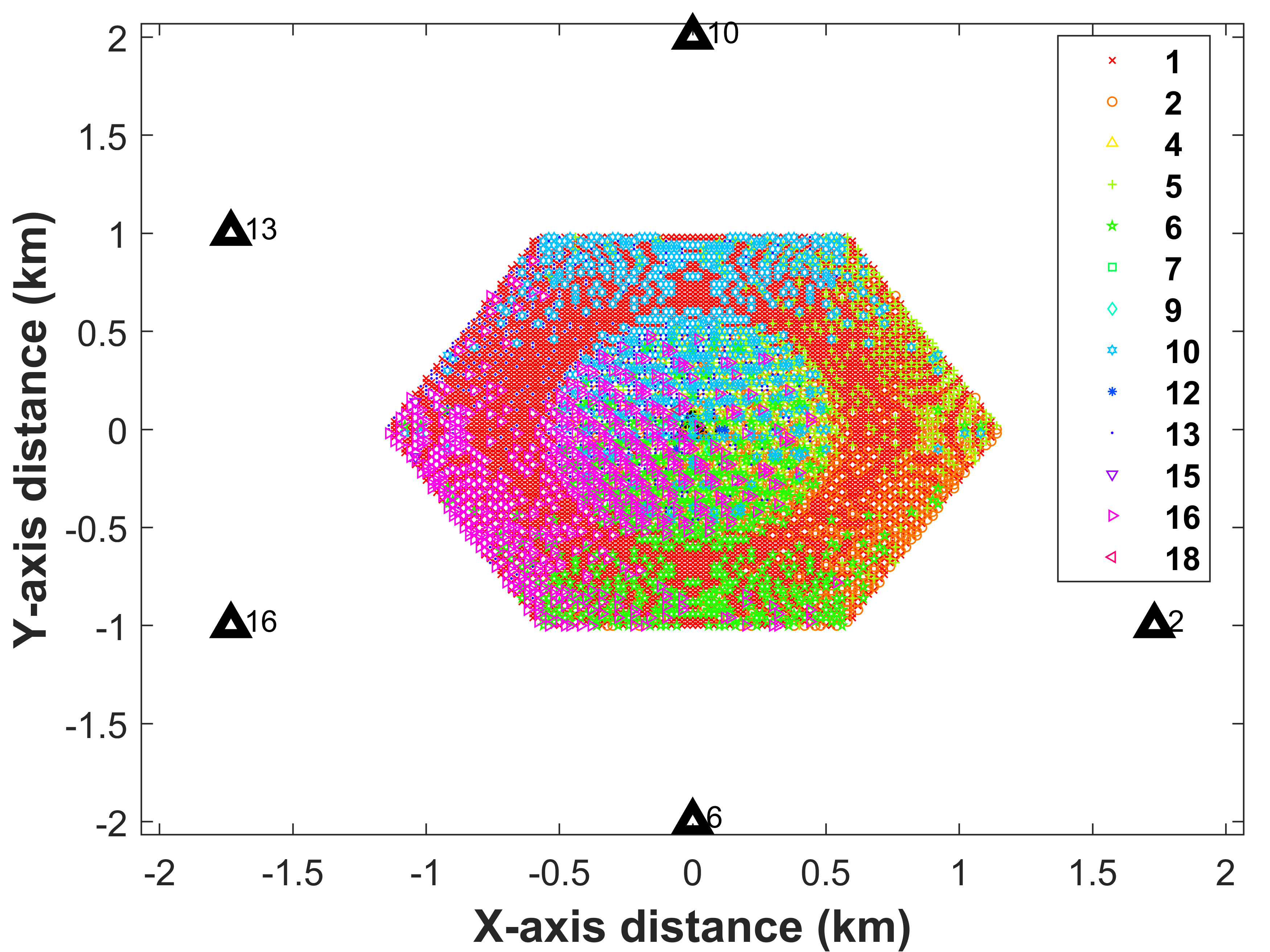}}
		\subfloat[ISD = $2000$~m, $h_{\textrm{UAV}}$ = $500$~m]{
			\includegraphics[width=.25\linewidth]{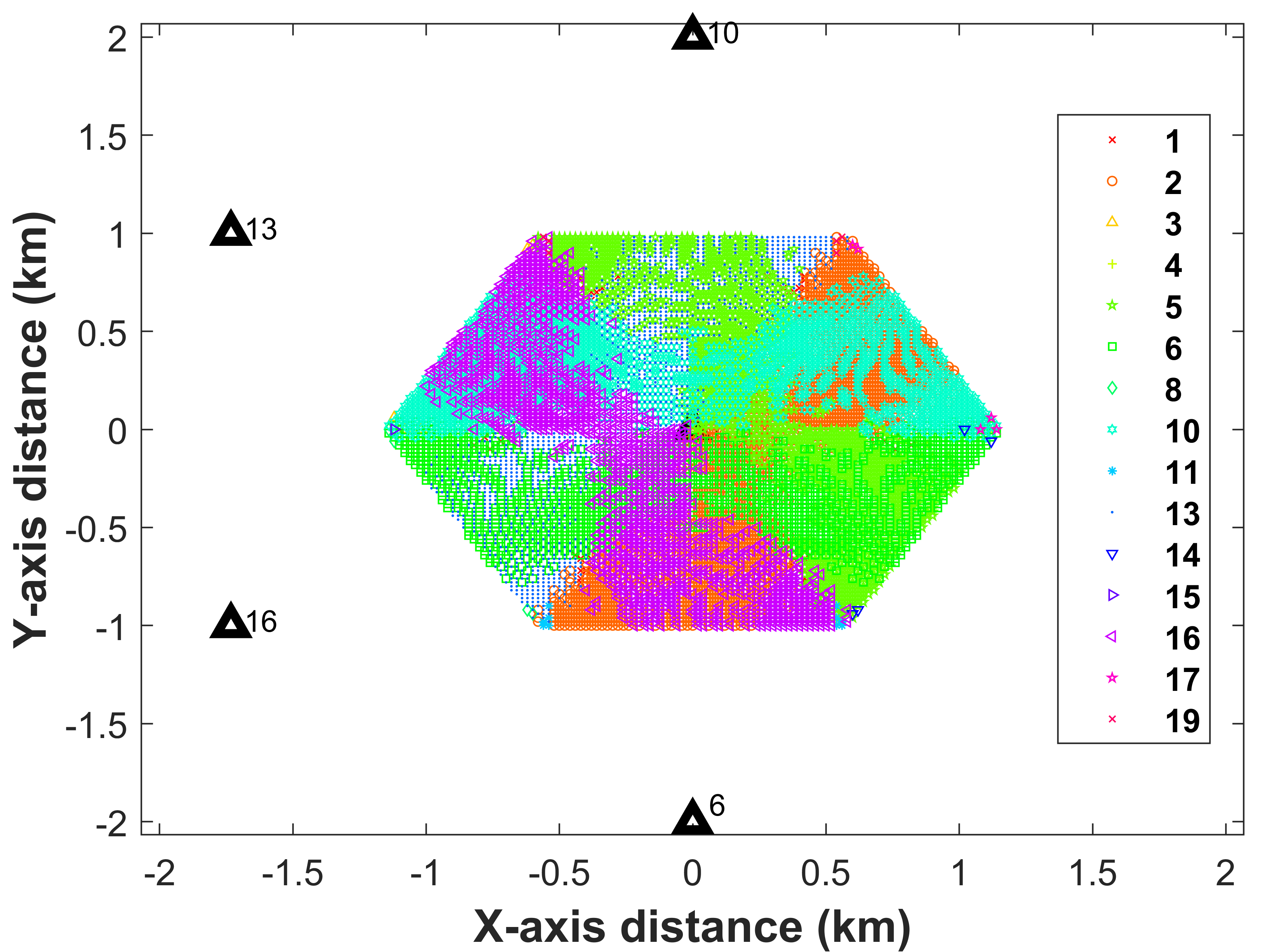}} \hfill
		\subfloat[ISD = $1000$~m, $h_{\textrm{UAV}}$ = $100$~m]{
		\includegraphics[width=.25\linewidth]{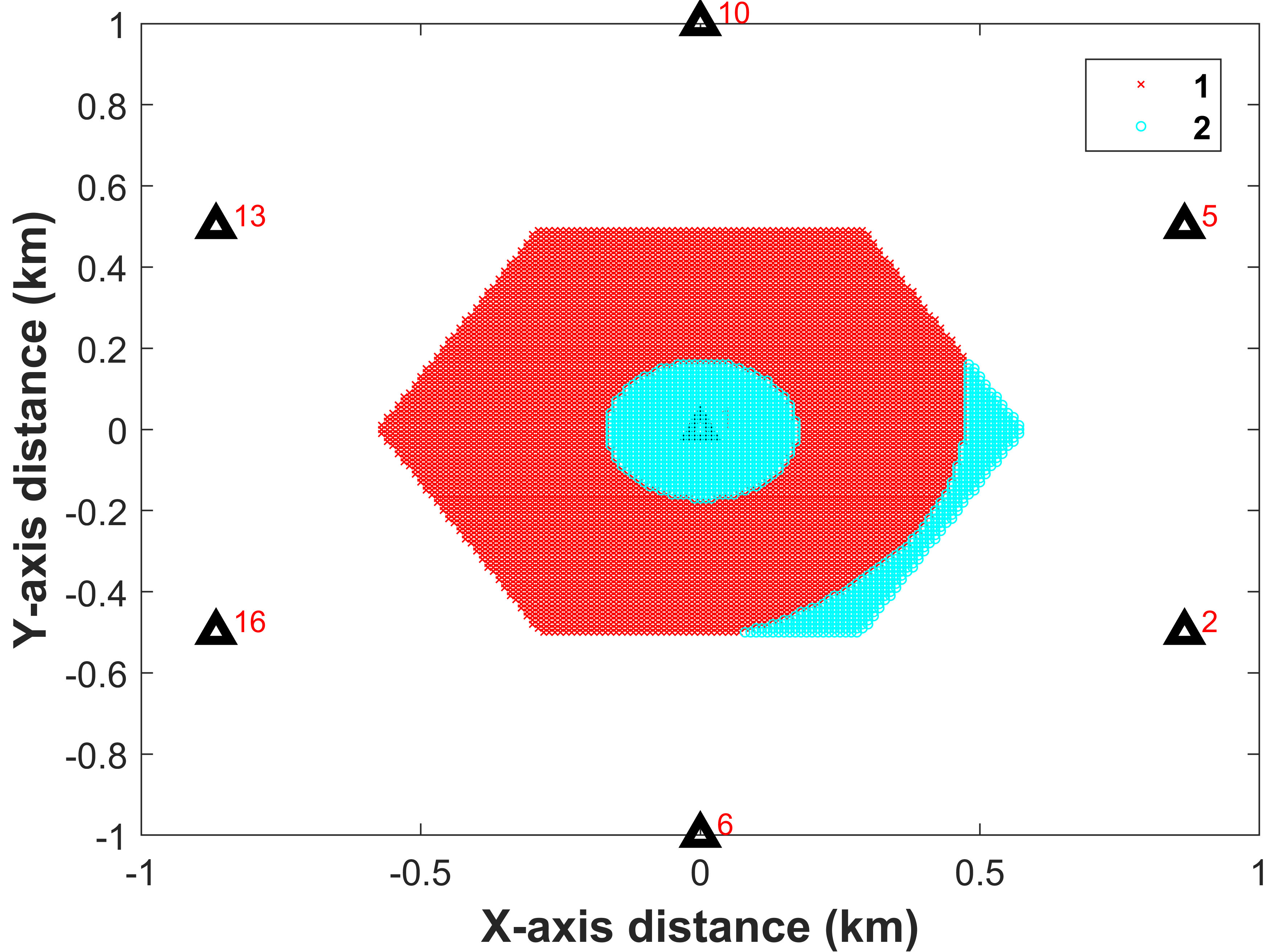}}
		\subfloat[ISD = $1000$~m, $h_{\textrm{UAV}}$ = $500$~m]{
			\includegraphics[width=.25\linewidth]{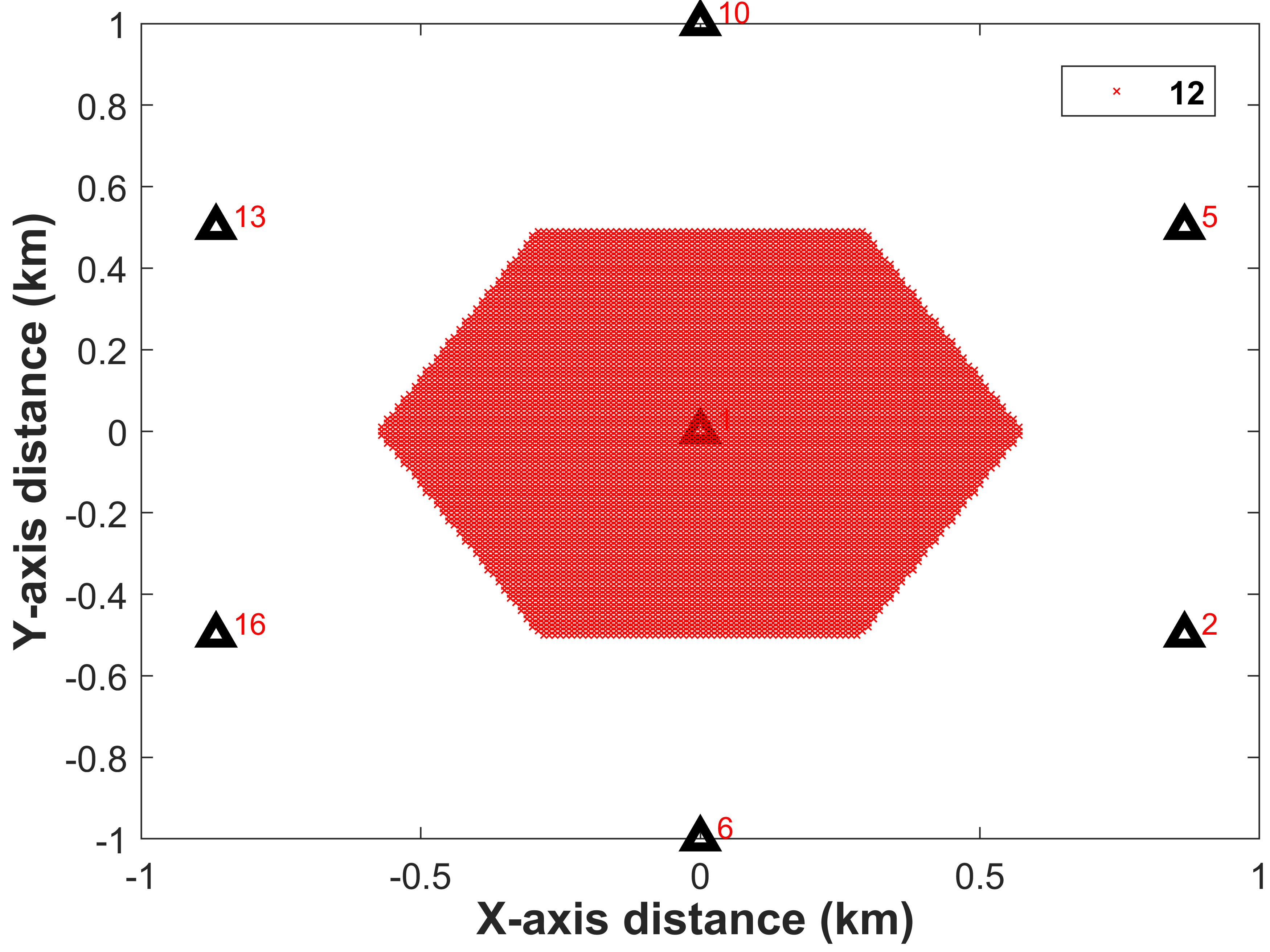}}
		\subfloat[ISD = $2000$~m, $h_{\textrm{UAV}}$ = $100$~m]{
		\includegraphics[width=.25\linewidth]{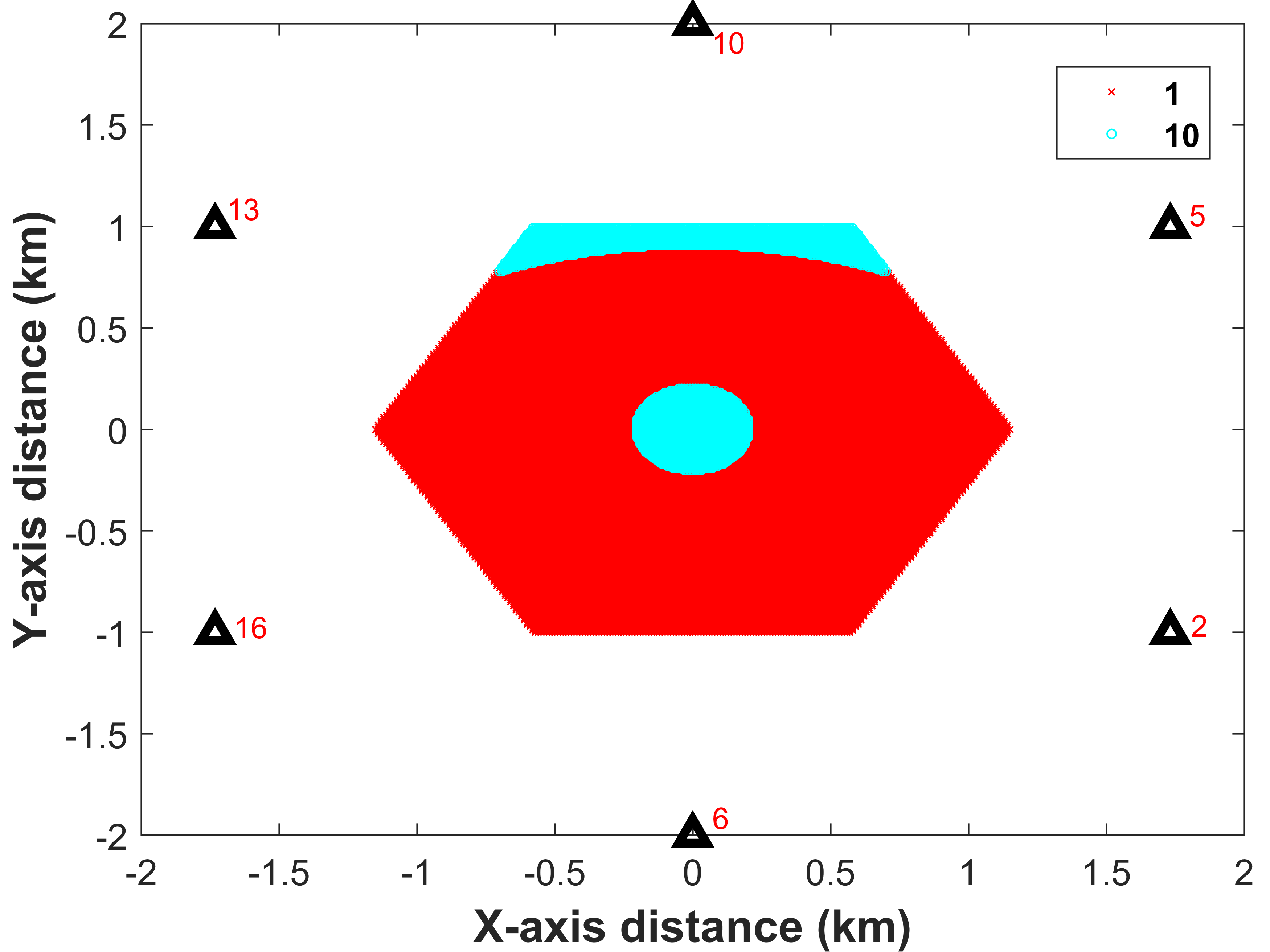}}
		\subfloat[ISD = $2000$~m, $h_{\textrm{UAV}}$ = $500$~m]{
			\includegraphics[width=.25\linewidth]{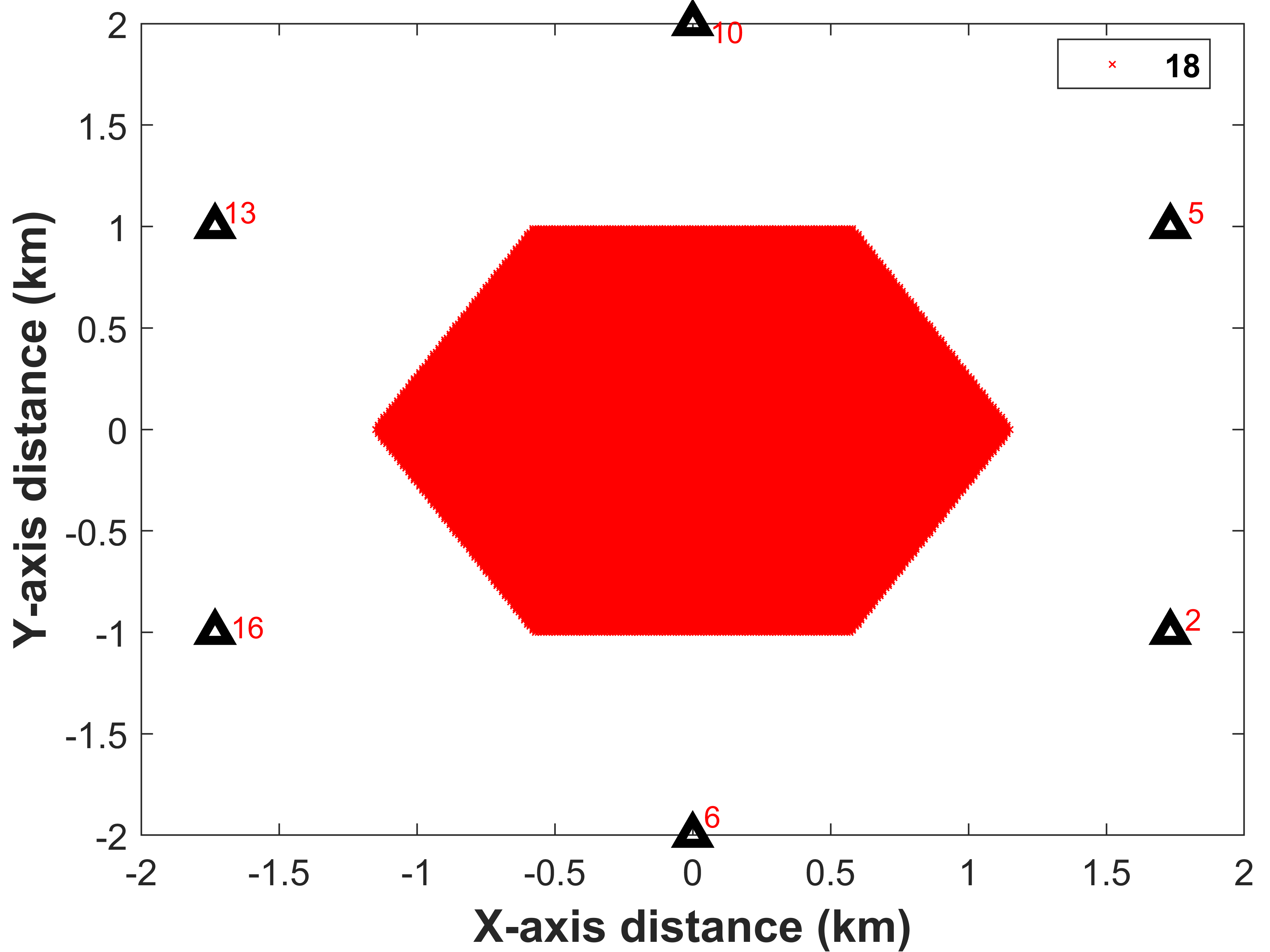}}
  \caption{{Cell association patterns for ISD $=1000$ and $2000$~m, and $h_{\textrm{UAV}}$ = $100$~m and $500$~m. Simulations are carried on using 19 cells, while only the coverage in the center cell is investigated. The top figures (a-d) represent the cases with downtilted antennas only. The bottom four figures (e-h) represent the hexagonal cells with optimized uptilted antennas. Overall, the patchy coverage due to the sidelobes of the downtilted antennas in (a-d) is mitigated in (e-h) due to the inclusion of the uptilted antennas with optimized uptilt angles that aim to maximize the minimum SIR.}}
   \label{fig:coverage}
\end{figure*}

\subsection{Potential Use of Cellular Networks for Air Corridors}
Due to their ubiquitous footprint, existing cellular networks have a strong potential to provide  C\&C connectivity to AVs. However, since the traditional cellular networks are optimized for ground users, their deployments bring major challenges to provide continuous and reliable coverage for aerial users. For instance, the main-lobes of the base station (BS) antennas are tilted toward the ground for providing better channel conditions for terrestrial users, and thus the UAVs flying over a specific altitude will be served by the sidelobes of the antennas. Therefore, either additional antennas or careful adjustment of existing antennas are required for AAM use. 

Moreover, UAVs flying high in the sky will be able to establish line of sight (LOS) links with several BSs, and this situation is well known to increase the overall interference in the AG downlink. Indeed, several industrial experiments have reported on the presence of strong inter-cell interference from other BS, and connection to UAVs through antenna side lobes, which in turn creates abrupt signal fluctuations as the UAVs change their locations. Another non-trivial source of interference is the ground reflection (GR) from the down-tilted main lobes of the BSs for both sub-$6$ GHz and millimeter-wave (mmWave) bands~\cite{chowdhury2021ensuring}. These factors together create challenges for using cellular networks as the sole connectivity medium for large-scale UAS deployment. 


\subsection{Numerical Results}

We study the impact of antenna radiation pattern and GR on the association pattern of the UAVs flying inside the drone corridors for two different inter-site distances (ISDs) and UAV heights $h_{\textrm{UAV}}$. We first consider that a part of a linear drone corridor falls inside the central hexagon where the $19$ cellular BSs are distributed in a two-tier hexagonal grid with a fixed inter-site distance (ISD). Hereinafter, we will use the terms 'BS' and 'cell' interchangeably. To average out the impact of AV/UAV distribution, we divide the center cell into discrete grid points, and a UAV is placed on each grid point at a height $h_{\textrm{UAV}}$. Here, we do not consider wrap-around and thus, we only consider the central hexagonal cell to capture the impact of inter-BS interference from the neighboring BSs. By considering flat fading channels, highest reference signal received power (RSRP) based association (HRA), and hexagonal cells, we report our finding for ISD$=1000$ and $2000$~m, and $h_{\textrm{UAV}}=100$~m and $h_{\textrm{UAV}}=500$~m~\cite{chowdhury2021ensuring}. The carrier frequency and BS transmission power are considered to be $2$~GHz and $46$~dBm, respectively.

To overcome the inherent shortcomings of traditional cellular networks, we propose deploying extra sets of antennas that are uptilted to provide good and reliable connectivity to the AVs/UAVs. These extra uptilted antennas coexist with the traditional down-tilted antennas and use the same time and frequency resources. However, introducing these extra antennas will increase the interference in both terrestrial and airborne segments, and hence it is critical to optimize the tilting angles of the main-lobes of these antennas~\cite{chowdhury2021ensuring }.~By using genetic algorithm (GA), the uptilt angles of these antennas are optimized to maximize the minimum signal-to-interference ratio of all of the discrete points inside the central hexagonal cell. Note that we use the modified path-loss model introduced in~\cite{chowdhury2021ensuring} to simulate the impact of GR and antenna sidelobes. For simplicity, we assume that all the A2G links are LOS. The BSs have antenna arrays with $8 \times 1$ elements and the main-lobes are downtilted by an angle of $6^{\circ}$.

In Fig.~\ref{fig:coverage}(a-d), we report the cell association patterns of these discrete points, which can be the segments of a drone corridor. For simplicity, we only show the six adjacent BSs (tier-$1$ BSs) to the center hexagonal cell. We can conclude that the joint impact of the BS antenna radiation pattern, higher LOS probability, and GR makes the coverage pattern in the sky uneven and patchy. A UAV flying through the drone corridor will thus make frequent handovers which can decrease the reliability of the C\&C communication in the downlink. The unevenness of the coverage pattern is especially more evident at $h_{\textrm{UAV}}=100$~m where the impact of GR is stronger than at higher altitude $h_{\textrm{UAV}}=500$~m~\cite{chowdhury2021ensuring}. 

The presence of uptilted antennas with optimized tilting angles towards the UAVs makes the coverage area smoother and more consistent, as depicted in  Fig.~\ref{fig:coverage}(e-h). Such a coverage trend is suitable for UAS deployment and most importantly, can guarantee higher minimum signal-to-interference-ratio (SIR) for the AVs. For  $h_{\textrm{UAV}}=100$~m, some discrete points will be served by a nearby BS whereas, for $h_{\textrm{UAV}}=500$~m, all of the discrete points will be served by a tier-$2$ BS. This is due to the fact that at higher UAV altitudes, the tier-$2$ GBSs can provide larger SIR by choosing an angle that covers most of the discrete UAV locations. The 3GPP specified sub-frame blanking (fully/partially) technique can also be introduced to provide even better connectivity for the UAS system, which will however come at a cost of fewer resources for the users associated with the downtilted antennas~\cite{chowdhury2021ensuring}. It is worth noting that the presence of these extra uptilted antennas will decrease the overall energy efficiency of the network and hence, a joint optimization of uptilt angles, transmit power of the uptilted antennas, and the downtilt angles of the downtilted antennas are required to guarantee the successful integration of UAS and cellular networks.

\subsection{Future Research Directions}

The research community is looking actively into providing reliable and safe connectivity for large-scale UAS deployment~\cite{facom,geraci2021}. The coverage using cellular networks can be improved using dedicated BSs or antennas for AVs, mmWave bands, dedicated resources for UAS, intelligent and robust beam tracking, among other approaches~\cite{facom}. Another wireless technology known as intelligent reflecting surface (IRS) can also be deployed on rooftops or building walls to increase the coverage in the sky in an energy-efficient manner~\cite{geraci2021}. However, for large-scale reliable UAS deployment, extensive measurement campaigns will be necessary to test the real-time applicability of these technologies. Cellular networks augmented with other resources, such as dedicated AAM ground stations, may also be worth investigating. Artificial intelligence (AI) or data-driven methods can be used to predict the coverage or association pattern in the drone corridors given the network and geographic topology nearby. Interested readers may refer to \cite{facom} and \cite{geraci2021} for further discussions.

\section{Mobility and Handovers for Aerial vehicles}

Since some AVs will need to travel long distances beyond visual line of sight (BVLOS), when  they are served by a cellular network, they will need to perform handover (HO) from one BS to another for seamless connectivity. Each HO requires transfer of buffered packets from one BS to another (for seamless handovers), and hence generates additional traffic in the core network. Unnecessary handovers hence need to be minimized while maintaining reliable network connectivity~\cite{6215543}. The number of handovers will depend on the altitude and velocity of the AV, which are studied in this section via computer simulations.

\subsection{Numerical Results}

\begin{figure}[t!]
    \begin{center}
        \includegraphics[width=0.9\linewidth]{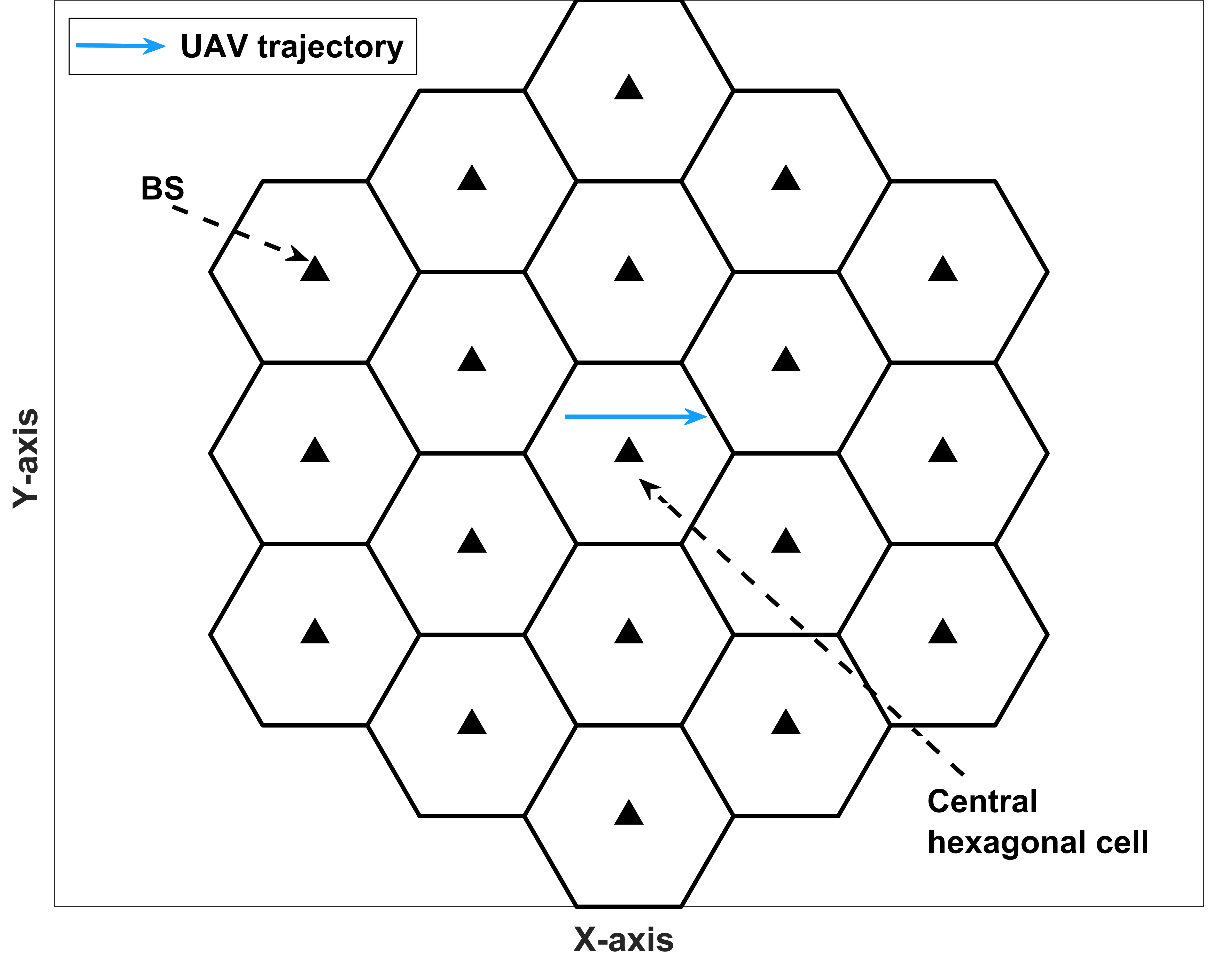}
        \caption{{Drone corridor inside the center cell of a two-tier cellular network.}}
    \label{fig:hexagon}
    \end{center}
    
\end{figure}

To study the mobility and handover performance of the AVs served by traditional cellular networks, we consider a similar type of hexagonal cellular network as presented in Section~III. Here, a single AV is flying along a two-dimensional (2D) linear trajectory (for instance, through the horizontal X-axis) at a fixed height $h_{\textrm{UAV}}$ and speed $v$ through a drone corridor. We consider the linear mobility model due to its simplicity and suitability for UAVs flying in the sky with virtually no obstacle inside the dedicated air corridors. A representation of such a setup is depicted in Fig.~\ref{fig:hexagon}. We also assume that once the AV reaches the boundary of the central hexagonal cell, it will ``bounce back" at a random angle. 

\begin{figure*}[t!]
\begin{center}
	\subfloat[ISD = $1000$~m, $h_{\textrm{UAV}}$ = $100$~m]{
			\includegraphics[width=0.45\linewidth]{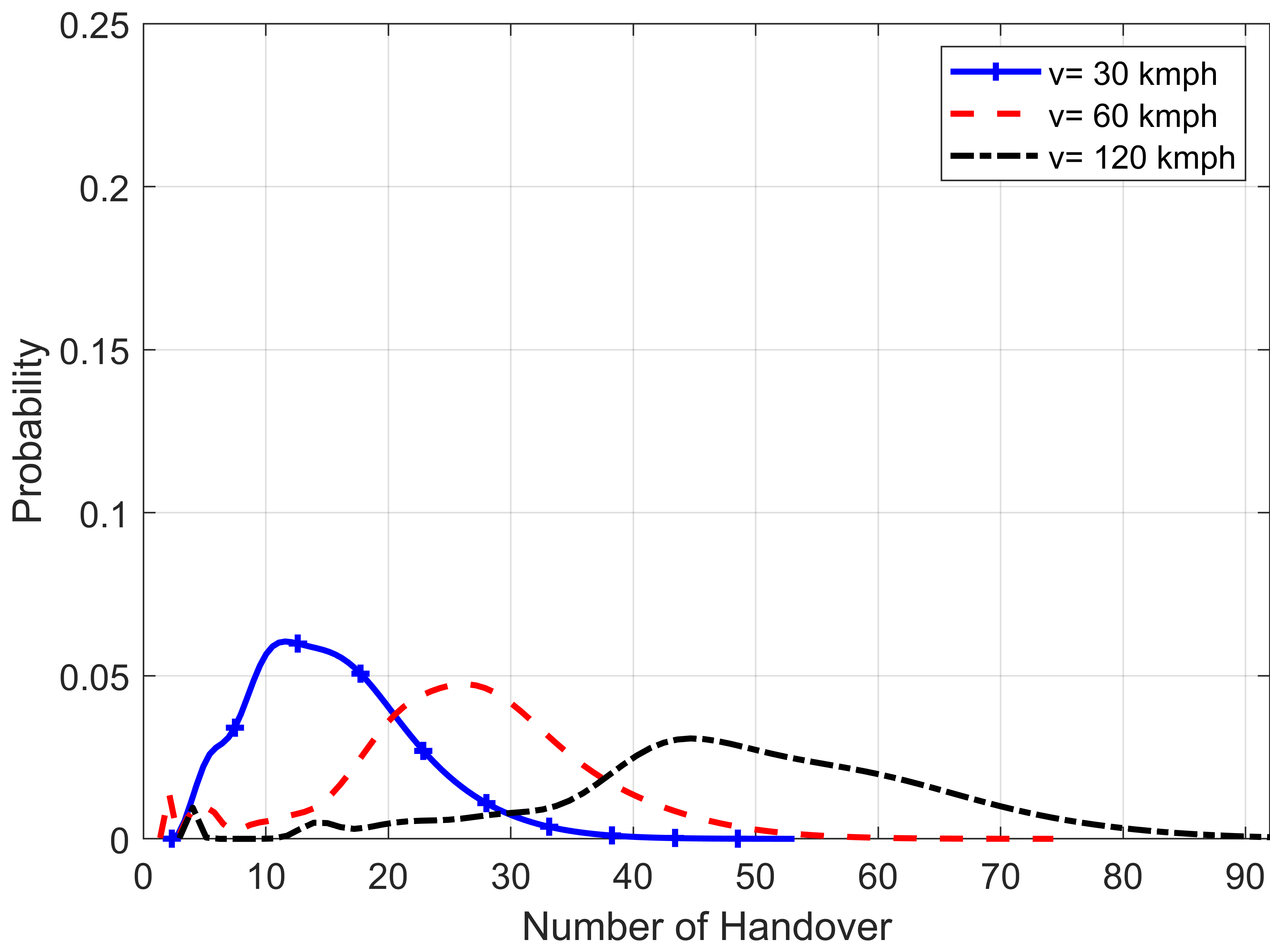}}
		\subfloat[ISD = $1000$~m, $h_{\textrm{UAV}}$ = $500$~m]{
			\includegraphics[width=0.45\linewidth]{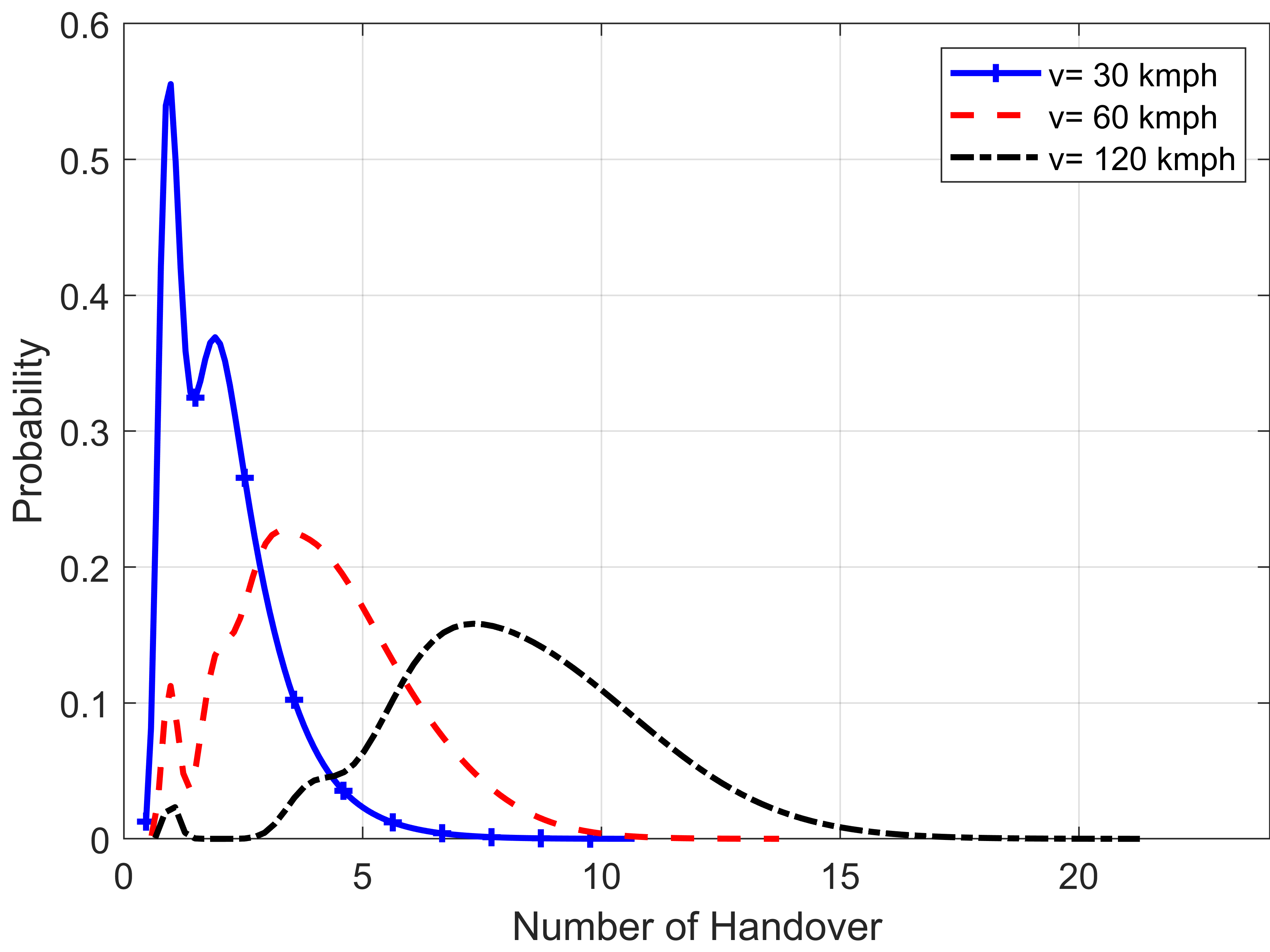}} \\
			\subfloat[ISD = $2000$~m, $h_{\textrm{UAV}}$ = $100$~m]{
			\includegraphics[width=0.45\linewidth]{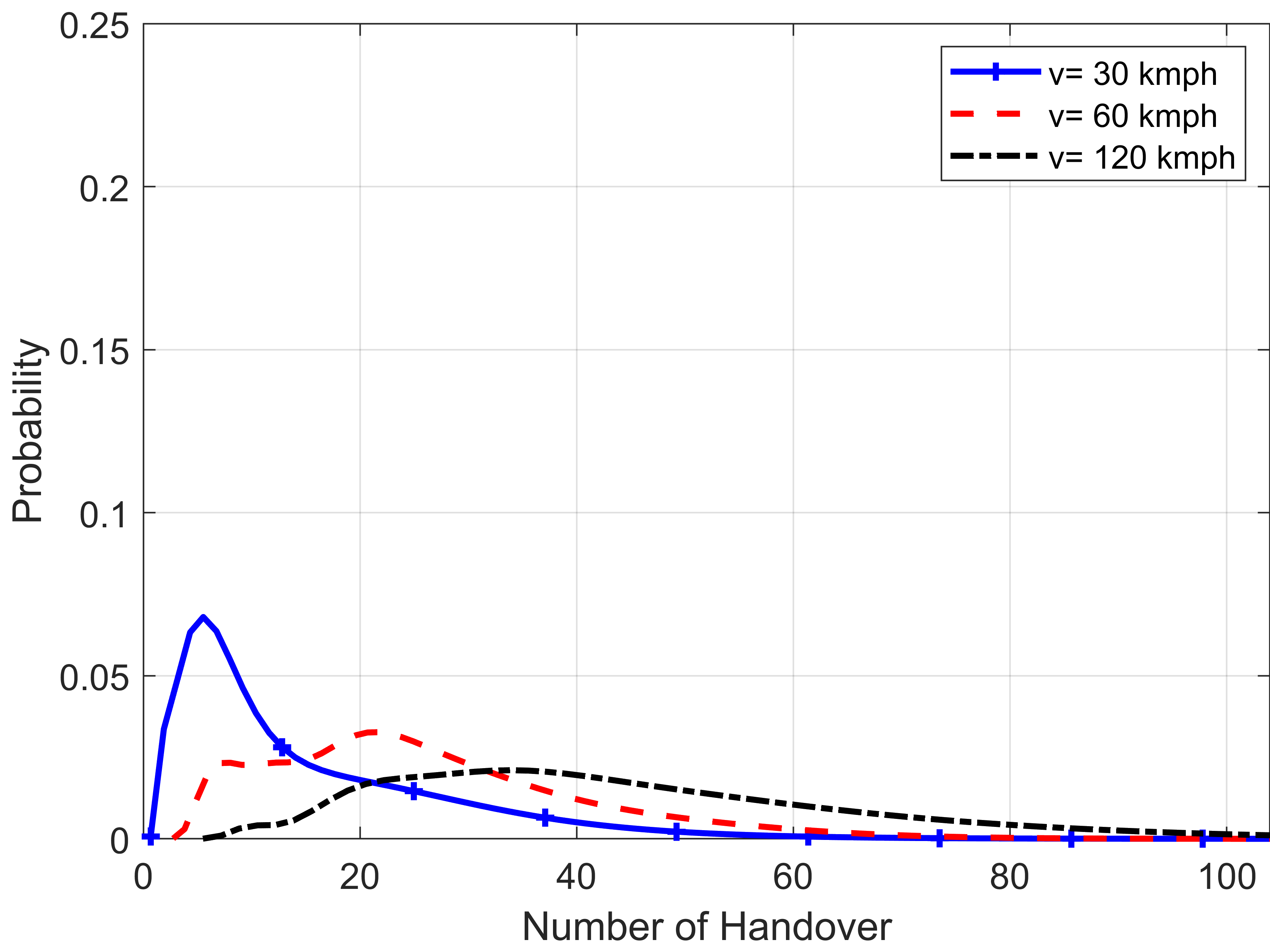}} 
			\subfloat[ISD = $2000$~m, $h_{\textrm{UAV}}$ = $500$~m]{
			\includegraphics[width=0.45\linewidth]{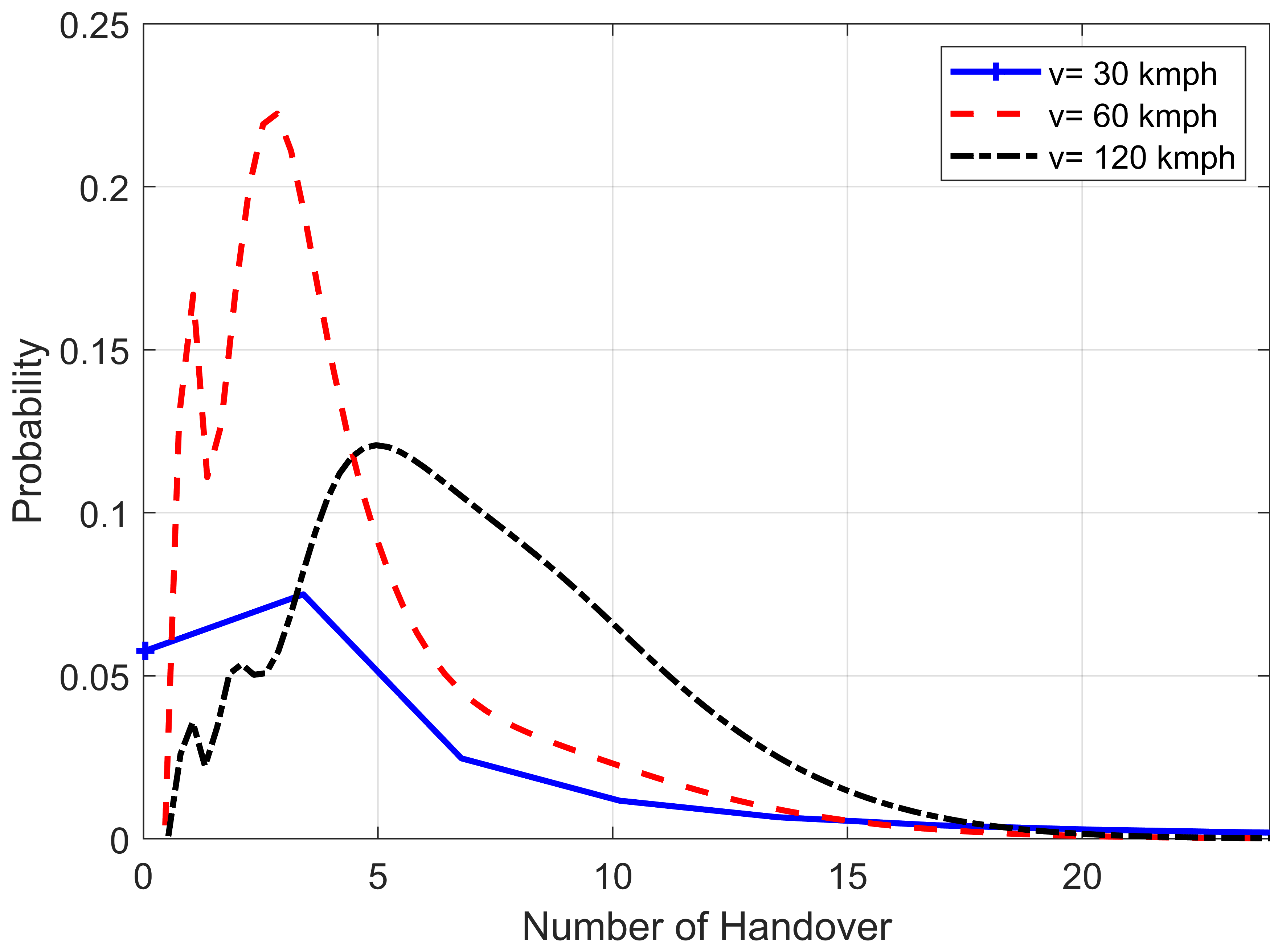}}
		\caption{{The PMFs of the HO count for different ISD and $h_{\textrm{UAV}}$ considering GR, correlated shadow fading, and antenna patterns.}}
		\label{fig:HO_count}
		 \end{center}
\end{figure*}

While flying, we consider that the network can track the number of HOs $H$ made by the AV during a measurement time window $T=3$~minutes. The HO procedure will follow the mechanism that involves a HO margin (HOM) parameter, and a time-to-trigger  (TTT)  parameter,  which is a time window that starts after the A3 event~\cite{3gpp,6215543}. Moreover, the shadow fading (SF) experienced by the AVs flying in higher altitudes will have strong correlations among the subsequent waypoints due to the high probability of LOS in the AG links~\cite{moin_ho}. 
Hence, we study the HO count performance of the cellular-connected AVs for different ISDs and $h_{\textrm{UAV}}$ combinations, and for random AV trajectories inside the center hexagonal cell. The pertinent results are reported in Fig.~\ref{fig:HO_count}. Here, TTT and HOM parameters are considered to be $40$~ms and $2$~dB, respectively. By considering SF as a first-order auto-regressive process, the auto-correlation parameter between the SF values at two points is chosen to be $0.82$~\cite{moin_ho}.

In Fig.~\ref{fig:HO_count}, we also consider three AV speeds, namely $30$, $60$, and $120$ kmph, and show the associated HO count probability mass functions (PMFs). We can conclude that overall, the AV makes fewer HOs for $h_{\textrm{UAV}}=500$~m. This is because at $h_{\textrm{UAV}}=500$~m, coverage becomes smoother than its $100$~m counterpart due to lower impact of the GR. However, the AVs will obtain weaker received signals due to the higher path-loss at high altitudes. Similar findings were also reported in~\cite{3dbeam_HO} where the authors studied the HO performance of cellular-connected drones by using beamforming and tracking in mmWave bands. Moreover, higher AV speeds usually result in higher HO counts due to the larger distances traveled by the AVs.

\subsection{Future Research Directions}

The mobility-related performance metrics such as HO count, HO failure, radio link failure, ping-pong HO, and minimum SIRs should be met with high reliability for city-wide UAS deployments. Though existing cellular networks may provide reliable mobility to a small number of AVs~\cite{3gpp}, for nationwide UAS deployment, some recent advances of 5G technologies such as massive MIMO with digital beamforming, wider bandwidths, and lower latency, can play major roles. For instance, the frequent HO and ``ping-pong" HO issues with downtilted cellular networks can be mitigated by using directional beams both at the BSs and at the AVs~\cite{geraci2021,3dbeam_HO}. As an example, ride-sharing company Uber has started working with AT\&T to support small piloted air-crafts at low altitudes by using 5G~\cite{geraci2021}. Supporting cellular-connected AVs with beyond 5G and the upcoming 6G technologies need to be studied further~\cite{geraci2021,Mozaffari2021Toward6W}.  

Apart from these, the HO-related parameters such as TTT, HOM, and measurement gaps can be optimized by using AI-based techniques to meet the mobility-related performance requirements. Moreover, since the trajectories of the AVs in the drone corridors are predictable and the location of the supporting BSs can be known in prior, AI can be used to associate AVs with the BSs for reliable mobility~\cite{Galkin2022REQIBARA}. A digital twin-based mobility and coverage study for different altitudes and BS deployments can be conducted before the initial deployment of AAM~\cite{digital_twin}. Research collaborations between the aviation, transportation, and wireless communication communities are of critical importance for the realization of safe and smart AAM.


\begin{figure*}
\begin{center}
    \includegraphics[width=0.8\linewidth]{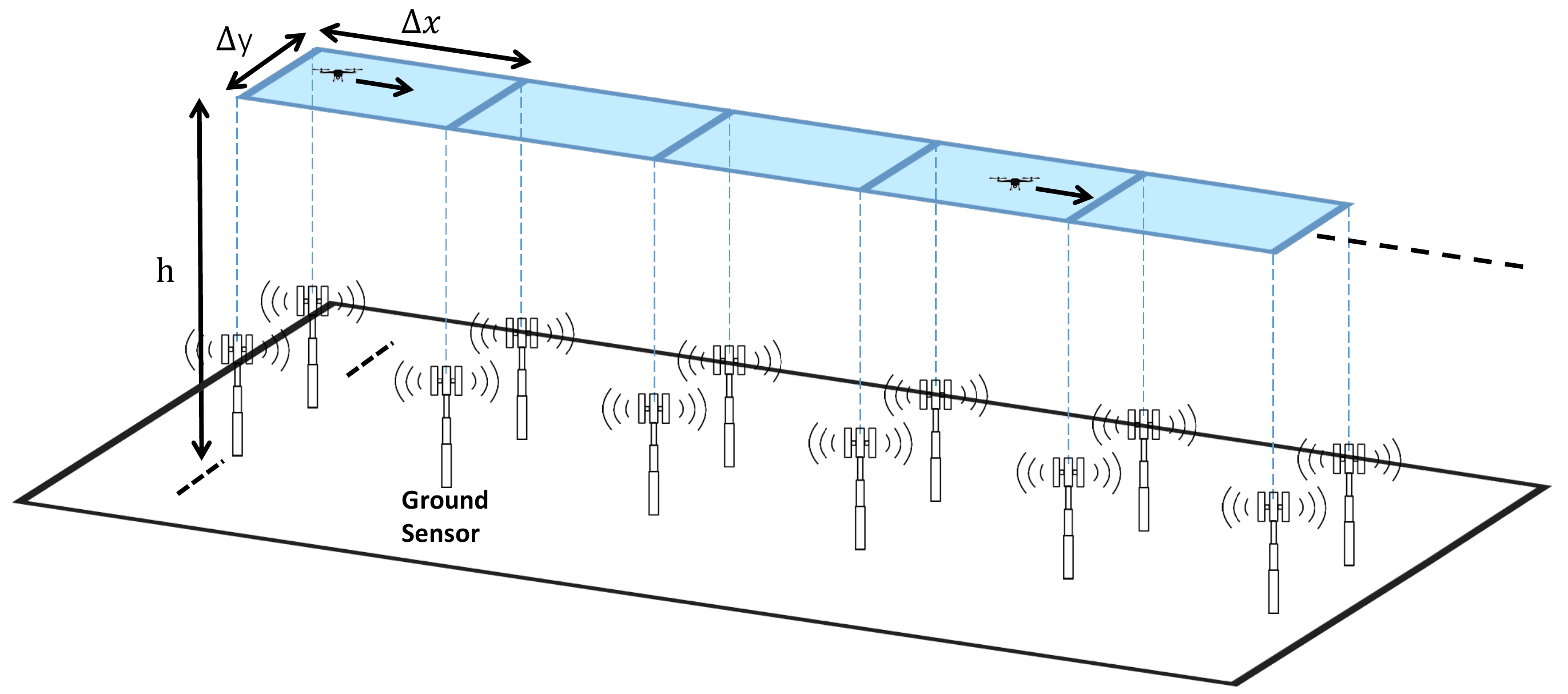}
\end{center}
\caption{Drone corridor scenario assumed for localization simulations. The localization accuracy is evaluated within each corridor segment at height $h$ with dimensions $\Delta x$ and $\Delta y$.}\label{Fig:CorridorLevels}
\end{figure*}

\section{Wireless Localization for Aerial Vehicles}
Safe and effective navigation of multiple UAVs is one of the most imperative and challenging problems that needs to be solved before drone corridors can be fully functional.  Such coordinated navigation in the drone corridor is crucial to preventing collisions, and managing fast and effective traffic flow in the corridor. This requires continuous exchange of current locations and future maneuvers between all UAVs involved. Moreover, when dealing with high speed UAVs, even a relatively small error in location estimates can lead to potential hazards. Thus, great care and attention must be paid to improving 3D localization accuracy in the drone corridor.

\begin{figure}[ht]
\centering
	\subfloat[$h_{\textrm{UAV}}$ = $100$~m]{
			\includegraphics[width=.9\linewidth]{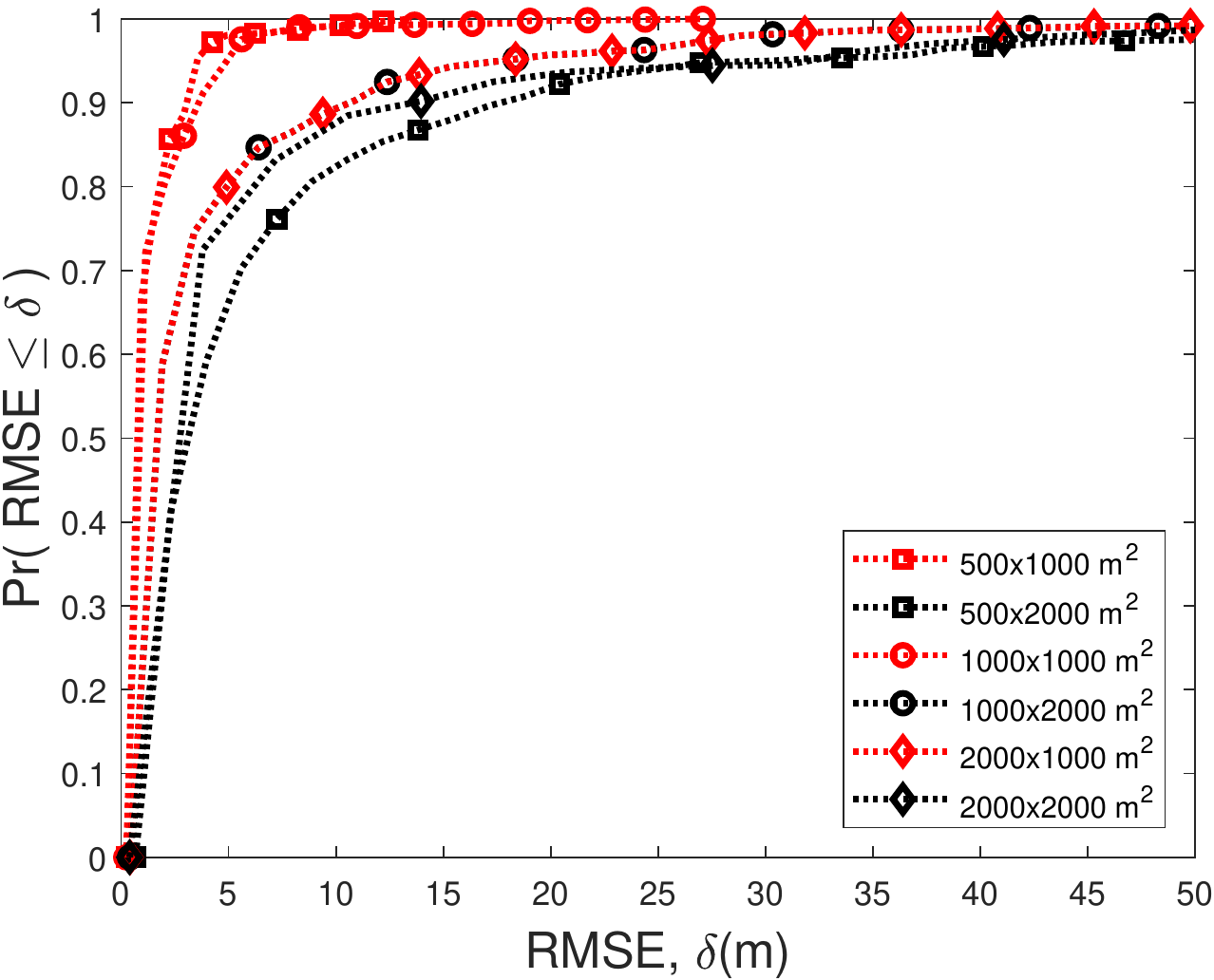}}\\
		\subfloat[$h_{\textrm{UAV}}$ = $500$~m]{
			\includegraphics[width=.9\linewidth]{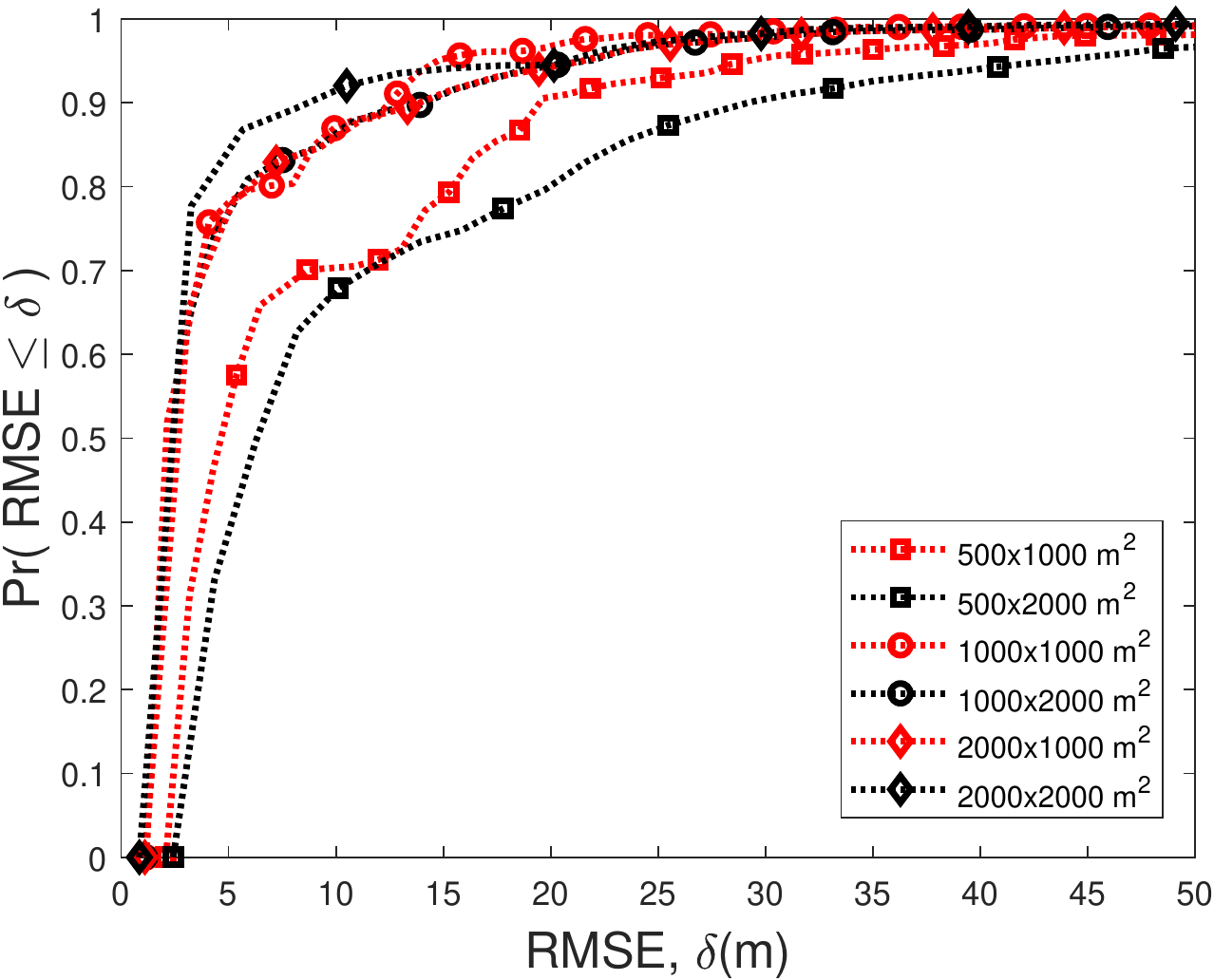}} 
			
  \caption{\textcolor{black}{CRLB RMSE distribution for multiple orthogonal antennas at UAV altitudes of (a) 100 meters, and (b) 500 meters.}}
\label{fig:crlb}
    \vspace{-2mm}
\end{figure}


There are several technical challenges to accurate 3D localization of UAVs that rely on measurements obtained from wireless communication links. Since
AG links in outdoor drone corridors have high likelihood of being a LOS link, the link quality is also highly sensitive to any possible mismatches in the transmit and the receive antenna patterns. As a result, the signal to noise ratio (SNR) of an AG link can degrade rapidly with slight changes in the transmit and receive antenna orientations and the mobility of the UAV. Since the quality of the signal measurements depends on the SNR of the AG link, it is important to take the 3D antenna patterns into account when designing a robust 3D localization schemes for the drone corridors. Due to their high accuracy and simplicity of implementation, time-based localization methods are a natural choice of RF parameter to be used for localization in a drone corridor.

\subsection{Multi-Antenna Techniques for Improved 3D Localization}

In order to study the effects of 3D antenna patterns on time-based localization in the drone corridor context,  we adopt the analytical framework put forth in \cite{92291678}. In \cite{92291678}, we consider a scenario where a fixed number of RF sensors equipped with single or multiple dipole antennas are placed at some known locations on the ground, and they derive the time-difference-of-arrival (TDOA) measurements from the time-of-arrival (TOA) data collected for the UAV that is also equipped with a dipole antenna. These measurements are then used to derive the Cramer-Rao lower bounds (CRLBs) on 3D location of the UAV, the localization error for various orientations of the dipole antennas at the transmitter and the receiver. Finally, \cite{92291678} quantifies the performance loss due to the possible antenna mismatches. 

In this present paper, we extend this framework to propose a mitigation technique, where each of the ground sensors and the UAV are equipped with two orthogonally oriented dipole antennas. We assume that one of these antennas is oriented along the vertical $z$-axis, and the other one along the horizontal $y$-axis. This allows us to leverage the antenna pattern diversity by transmitting two different versions of the same signal and then by receiving two different copies of the same composite received signal via two receive antennas with two different orientations. The two copies of the received signal from the two receive antennas are then combined using the principles of maximal ratio combining (MRC). Although \cite{92291678} considers only a $3$D multiple input single output (MISO) link, where the transmitter is equipped with only one antenna but the receiver is equipped with two orthogonal dipole antennas, in this paper we consider a $3$D multiple input multiple output (MIMO) link, where both the transmitter and the receiver are equipped with two orthogonal dipole antennas.

Although the multi-antenna techniques are bound to give better localization performance, when compared to the single antenna techniques, the dimension of the corridor and the particular locations of the ground sensors in the corridor together determine the total antenna gain experienced by A2G links in the corridor. Thus, in order to design a corridor where UAV localization is robust to the variations in the patterns arising from the high mobility of the UAVs, it is important to understand the localization error as a function of the corridor dimension. Towards this end we derive the CRLB for the above mentioned multi-antenna scenario, following the steps shown in \cite{92291678}. 

\subsection{Numerical Results}

\begin{figure}[t!]
    \begin{center}
        \includegraphics[width=0.9\linewidth]{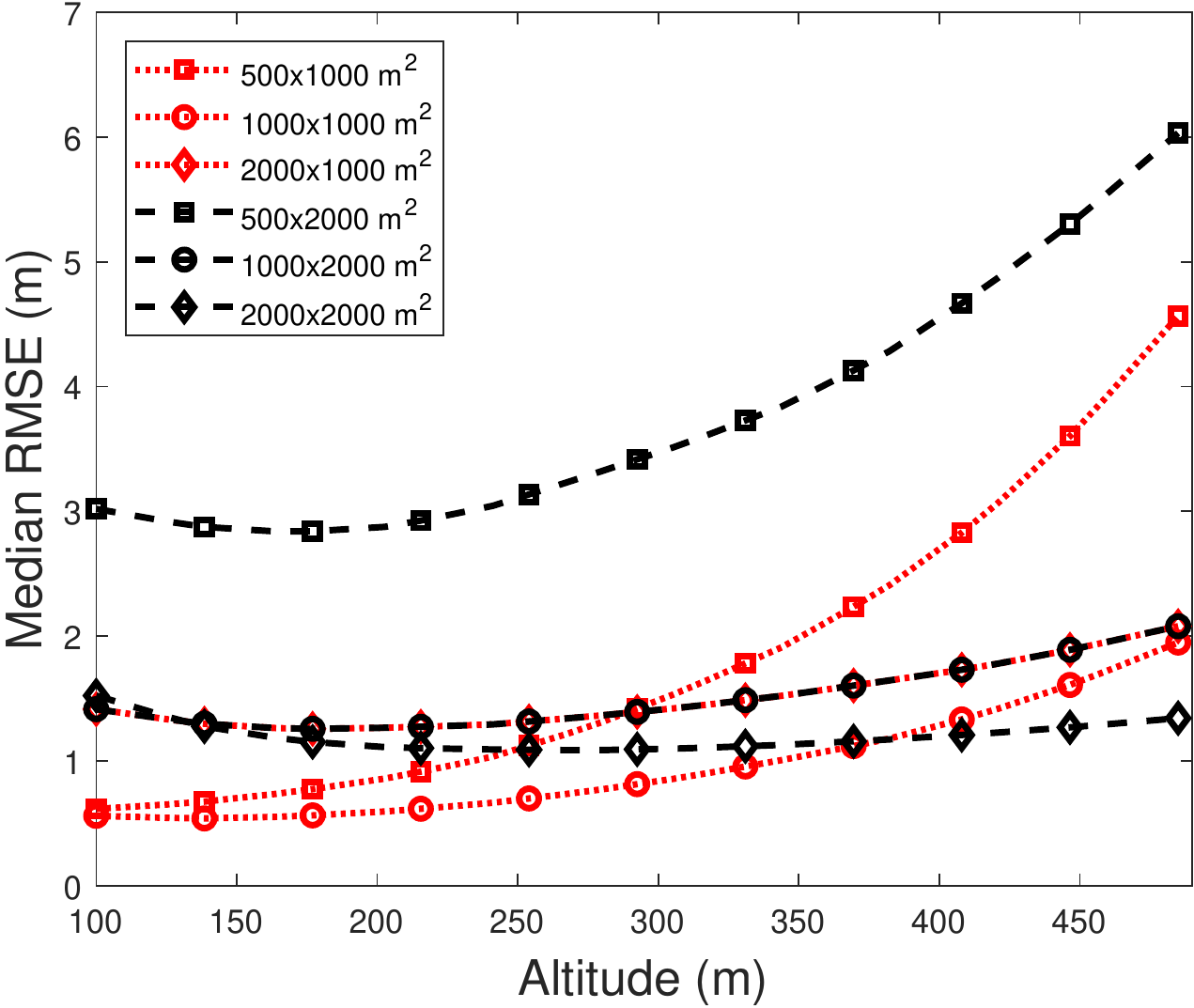}
        \caption{{\textcolor{black}{Median RMSE as a function of the UAV altitude for various different corridor dimensions}.}}
    \label{fig:med}
    \end{center}
    
\end{figure}

In Fig.~\ref{fig:crlb} and Fig.~\ref{fig:med}, we plot the derived localization CRLB for different corridor dimensions and UAV altitudes. In particular, we consider a slice of the corridor of size $\Delta x{\times}\Delta y\,{\text{m}^2}$ with $4$ RF sensors placed at the following 2D coordinates: $\textcolor{black}{\boldsymbol{x_1}}=(\frac{\Delta x}{2},\frac{\Delta y}{2})\,{\text{m}}$, $\textcolor{black}{\boldsymbol{x_2}}=({-}\frac{\Delta x}{2},\frac{\Delta y}{2})\,{\text{m}}$, $\textcolor{black}{\boldsymbol{x_3}}=({-}\frac{\Delta x}{2},{-}\frac{\Delta y}{2})\,{\text{m}}$, and $\textcolor{black}{\boldsymbol{x_4}=(\frac{\Delta x}{2},{-}\frac{\Delta y}{2})\,{\text{m}}}$. For the numerical analysis, we consider $\Delta x \in \{500, 1000, 2000\}$ meters and $\Delta y \in \{1000, 2000\}$ meters. We also assume that the transmit power of the UAV is $20$~dBm, and the bandwidth and operating frequency of the communication band are $10$~MHz, and $2.4$~GHz, respectively. Lastly, we describe the radiation pattern of the individual antennas as that of a half-wave dipole antenna.

As demonstrated by Fig.~\ref{fig:crlb}(a), we observe that at moderately low UAV altitudes, such as $100$ meters, the cumulative distribution functions (CDFs) for the narrower corridor configuration, i.e. $\Delta y=1000$ meters, lie above the CDFs for wider corridor configuration (i.e. $\Delta y=2000$), for all values of the localization RMSE, $\delta$. This implies that, at the very low altitudes, the best localization coverage $\text{Pr}({\text{RMSE}} \le \delta)$, \textcolor{black}{for the Model-1 pattern}, is provided by the narrower corridor. On the contrary, the curves in Fig.~\ref{fig:crlb}(b) show us that localization performance is largely dependent on the total area of the 3D space in consideration. We observe that  the highest coverage probability subject to very low error tolerances (such as $\delta \leq 20$ m), is obtained for the corridor with the largest area of $4$ SqKm, followed by the space with total area between $2$ SqKm to $1$ SqKm, and the space with total area between $1$ SqKm to $0.5$ SqKm. 

In Fig.~\ref{fig:med}, we characterize the median localization error as a function of the UAV altitude, for a narrow and a wide corridor, respectively. We note that among the 3D spaces with the wider corridor configuration, the ones with the smallest area of $0.5$ SqKm and $1$ SqKm experience the largest increase in localization error, as the drone altitude increases. However the localization error for other wider configurations with the larger areas remains almost invariant to the drone altitude.

\subsection{Future Research Directions}

As stated, safe navigation of UAVs in a drone corridor is completely dependent on accurate self-localization of UAVs or cooperative localization of a group of UAVs. In state of the art systems, such drift-free navigation is achieved by combining the inertial sensing of the UAVs with GPS. However such systems can fail in many situations, where reliable GPS connection is unavailable or vulnerable to outside attacks~\cite{ellis2020time}. For example, satellite coverage often breaks down in urban canyons or mountains due to high density of obstructions and the presence of rich multipath reflection and scattering. GPS systems can also be rendered ineffective by adversarial jamming and spoofing. Due to these reasons the GPS-denied navigation has gained a lot of attention, and the drone corridors provide an adequately complex infrastructure that requires in-depth research to devise GPS free localization schemes that are tailored to the drone corridor context.

In the context of cooperative localization, drones equipped with passive radars can also be considered in conjunction with a drone corridors. For problems like this, both classical and data-driven machine learning algorithms for fusion of the measurements from the ground sensors and radars mounted on the drones can be extremely beneficial to improving the localization accuracy. However, when relying on radar technologies in the drone corridor, intensive research must be carried out to differentiate between birds and small drones. Given the low weights and small radar cross sections, small drones can be easily confused with birds, creating unmanageable situations for operation of drone corridors. Such problems can get further exacerbated in certain areas such as seaside spots that are natural habitats for large populations of birds. Thus there is a great need for research and planning that adequately accounts for such events in a drone corridor context.   

The particular idea of using 3D localization CRLB for characterization of localization schemes can also be extended to more complicated antenna pattern scenarios, such as down-tilted sector antennas used by cellular BSs. Analogous to the approach presented in \cite{92291678}, expressions for the lower bounds on the localization error in a corridor setting with arbitrary ground sensor location can be derived and used to find the optimal location of the ground sensors with non-isotropic antenna patterns. 
Finally, there is a strong need in doing experiments over related UAS testbeds, such as the NSF AERPAW platform at NC State University~\cite{marojevic2020advanced}, to test and validate new concepts in practical air corridor scenarios. 
\section{Concluding Remarks}
This article underscores the importance of drone communications as the community moves towards realizing the full potential AAM services. It highlights that the direct  air-to-air communications among the AVs is the key enabler for safe and efficient flight operations in air corridors. Focusing on  cellular networks, we present various concepts and strategies for enhanced coverage, mobility management, and  wireless localization of  AVs in air corridors. Future research directions are also discussed in connection to each of these areas. 

\bibliographystyle{IEEEtran}
\bibliography{bibfiles/references}

\begin{thebibliography}{10}
\providecommand{\url}[1]{#1}
\csname url@samestyle\endcsname
\providecommand{\newblock}{\relax}
\providecommand{\bibinfo}[2]{#2}
\providecommand{\BIBentrySTDinterwordspacing}{\spaceskip=0pt\relax}
\providecommand{\BIBentryALTinterwordstretchfactor}{4}
\providecommand{\BIBentryALTinterwordspacing}{\spaceskip=\fontdimen2\font plus
\BIBentryALTinterwordstretchfactor\fontdimen3\font minus
  \fontdimen4\font\relax}
\providecommand{\BIBforeignlanguage}[2]{{%
\expandafter\ifx\csname l@#1\endcsname\relax
\typeout{** WARNING: IEEEtran.bst: No hyphenation pattern has been}%
\typeout{** loaded for the language `#1'. Using the pattern for}%
\typeout{** the default language instead.}%
\else
\language=\csname l@#1\endcsname
\fi
#2}}
\providecommand{\BIBdecl}{\relax}
\BIBdecl

\bibitem{NASA_AAM_Vision}
\BIBentryALTinterwordspacing
{NASA}, ``Advanced air mobility mission overview.'' [Online]. Available:
  \url{https://www.nasa.gov/aam}
\BIBentrySTDinterwordspacing

\bibitem{bauranov2021designing}
A.~Bauranov and J.~Rakas, ``Designing airspace for urban air mobility: A review
  of concepts and approaches,'' \emph{Progress in Aerospace Sciences}, vol.
  125, p. 100726, 2021.

\bibitem{lin2018sky}
X.~Lin, V.~Yajnanarayana, S.~D. Muruganathan, S.~Gao, H.~Asplund, H.-L.
  Maattanen, M.~Bergstrom, S.~Euler, and Y.-P.~E. Wang, ``The sky is not the
  limit: {LTE} for unmanned aerial vehicles,'' \emph{IEEE Commun. Mag.},
  vol.~56, no.~4, pp. 204--210, 2018.

\bibitem{chowdhury2021ensuring}
M.~M.~U. {Chowdhury}, I.~{Guvenc}, W.~{Saad}, and A.~{Bhuyan}, ``{Ensuring
  Reliable Connectivity to Cellular-Connected {UAV}s with Uptilted Antennas and
  Interference Coordination},'' \emph{ITU J. Future and Evolving Technol. (ITU
  JFET)}, Dec. 2021.

\bibitem{amorim2018measured}
R.~Amorim, H.~Nguyen, J.~Wigard, I.~Z. Kov{\'a}cs, T.~B. S{\o}rensen, D.~Z.
  Biro, M.~S{\o}rensen, and P.~Mogensen, ``Measured uplink interference caused
  by aerial vehicles in {LTE} cellular networks,'' \emph{IEEE Wireless Commun.
  Lett.}, vol.~7, no.~6, pp. 958--961, 2018.

\bibitem{al2017modeling}
A.~Al-Hourani and K.~Gomez, ``Modeling cellular-to-{UAV} path-loss for suburban
  environments,'' \emph{IEEE Wireless Commun. Lett.}, vol.~7, no.~1, pp.
  82--85, 2017.

\bibitem{muna2021air}
S.~I. Muna, S.~Mukherjee, K.~Namuduri, M.~Compere, M.~I. Akbas, P.~Moln{\'a}r,
  and R.~Subramanian, ``Air corridors: Concept, design, simulation, and rules
  of engagement,'' \emph{Sensors}, vol.~21, no.~22, p. 7536, 2021.

\bibitem{ellis2020time}
K.~Ellis, J.~Koelling, M.~Davies, and P.~Krois, ``In-time system-wide safety
  assurance ({ISSA}) concept of operations and design considerations for urban
  air mobility ({UAM}),'' \emph{NASA/TM-2020-5003981}, June 2020.

\bibitem{8741719}
N.~Hosseini, H.~Jamal, J.~Haque, T.~Magesacher, and D.~W. Matolak, ``{UAV}
  command and control, navigation and surveillance: A review of potential {5G}
  and satellite systems,'' in \emph{Proc. IEEE Aerospace Conf.}, Big Sky, MO,
  2019, pp. 1--10.

\bibitem{facom}
A.~Baltaci, E.~Dinc, M.~Ozger, A.~Alabbasi, C.~Cavdar, and D.~Schupke, ``A
  survey of wireless networks for future aerial communications (facom),''
  \emph{IEEE Commun. Surveys Tuts.}, vol.~23, no.~4, pp. 2833--2884, 2021.

\bibitem{3gpp}
\BIBentryALTinterwordspacing
3GPP, Technical Specification (TS) 36.777, 2018. [Online]. Available:
  \url{https://portal.3gpp.org/desktopmodules/Specifications/\\SpecificationDetails.aspx?specificationId=3231}
\BIBentrySTDinterwordspacing

\bibitem{geraci2021}
G.~{Geraci}, A.~{Garcia-Rodriguez}, M.~{Mahdi Azari}, A.~{Lozano},
  M.~{Mezzavilla}, S.~{Chatzinotas}, Y.~{Chen}, S.~{Rangan}, and M.~{Di Renzo},
  ``What will the future of {UAV} cellular communications be? {A} flight from
  {5G to 6G},'' \emph{arXiv e-prints}, p. arXiv:2105.04842, May 2021.

\bibitem{6215543}
D.~López-Pérez, I.~Güvenç, and X.~Chu, ``Mobility enhancements for
  heterogeneous networks through interference coordination,'' in \emph{Proc.
  IEEE Wireless Commun. Netw. Conf. Workshops (WCNCW)}, 2012, pp. 69--74.

\bibitem{moin_ho}
M.~M.~U. Chowdhury, P.~Sinha, and I.~Güvenç, ``Handover-count based velocity
  estimation of cellular-connected {UAV}s,'' in \emph{Proc. IEEE SPAWC},
  Atlanta, GA, 2020, pp. 1--5.

\bibitem{3dbeam_HO}
A.~Colpaert, E.~Vinogradov, and S.~Pollin, ``3{D} beamforming and handover
  analysis for {UAV} networks,'' in \emph{Proc. IEEE Globecom Workshops (GC
  Wkshps}, 2020, pp. 1--6.

\bibitem{Mozaffari2021Toward6W}
M.~Mozaffari, X.~Lin, and S.~Hayes, ``Toward {6G} with connected sky: {UAV}s
  and beyond,'' \emph{IEEE Commun. Mag.}, vol.~59, pp. 74--80, 2021.

\bibitem{Galkin2022REQIBARA}
B.~Galkin, E.~Fonseca, R.~Amer, L.~A. DaSilva, and I.~Dusparic, ``Reqiba:
  Regression and deep {Q}-learning for intelligent {UAV} cellular user to base
  station association,'' \emph{IEEE Trans. Veh. Technol.}, vol.~71, pp. 5--20,
  2022.

\bibitem{digital_twin}
H.~X. Nguyen, R.~Trestian, D.~To, and M.~Tatipamula, ``Digital twin for {5G}
  and beyond,'' \emph{IEEE Commun. Mag.}, vol.~59, no.~2, pp. 10--15, 2021.

\bibitem{92291678}
P.~Sinha and I.~Guvenc, ``Impact of antenna pattern on {TOA} based 3{D} {UAV}
  localization using a terrestrial sensor network,'' \emph{IEEE Trans. Veh.
  Technol.}, 2022.

\bibitem{marojevic2020advanced}
V.~Marojevic, I.~Guvenc, R.~Dutta, M.~L. Sichitiu, and B.~A. Floyd, ``Advanced
  wireless for unmanned aerial systems: {5G} standardization, research
  challenges, and {AERPAW} architecture,'' \emph{IEEE Veh. Technol. Mag.},
  vol.~15, no.~2, pp. 22--30, 2020.

\end{thebibliography}

\end{document}